\documentclass[conference,compsoc]{IEEEtran}
\usepackage{graphicx}
\usepackage[dvipsnames]{xcolor}
\usepackage{xurl}
\usepackage{listings}
\usepackage[inline]{enumitem}
\usepackage{textcomp}
\newcommand{\mytexttilde}{\raisebox{0.5ex}{\texttildelow}}
\usepackage{booktabs}
\usepackage{amssymb}
\usepackage{caption}
\usepackage[compact]{titlesec}
\titlespacing{\subsection}{0pt}{*1.5}{*0.7}
\ifCLASSOPTIONcompsoc
  \usepackage[nocompress]{cite}
\else
  \usepackage{cite}
\fi
\ifCLASSINFOpdf
\else
\fi
\begin{document}
\title{A Systematization of Security Vulnerabilities in Computer Use Agents}
\author{
\IEEEauthorblockN{
    Daniel Jones\thanks{Corresponding author: jonesdaniel@microsoft.com}, 
    Giorgio Severi, Martin Pouliot, Gary Lopez, Joris de Gruyter, Santiago Zanella-Beguelin,\\
    Justin Song, Blake Bullwinkel, Pamela Cortez, Amanda Minnich
}
\IEEEauthorblockA{Microsoft}
}
\maketitle
\begin{abstract}
Computer Use Agents (CUAs), autonomous systems that interact with software interfaces via browsers or virtual machines, are rapidly being deployed in consumer and enterprise environments. These agents introduce novel attack surfaces and trust boundaries that are not captured by traditional threat models. Despite their growing capabilities, the security boundaries of CUAs remain poorly understood.
In this paper, we conduct a systematic threat analysis and testing of real-world CUAs under adversarial conditions. We identify seven classes of risks unique to the CUA paradigm, and analyze three concrete exploit scenarios in depth: (1) clickjacking via visual overlays that mislead interface-level reasoning, (2) indirect prompt injection that enables Remote Code Execution (RCE) through chained tool use, and (3) CoT exposure attacks that manipulate implicit interface framing to hijack multi-step reasoning.
These case studies reveal deeper architectural flaws across current CUA implementations. Namely, a lack of input provenance tracking, weak interface-action binding, and insufficient control over agent memory and delegation. We conclude by proposing a CUA-specific security evaluation framework and design principles for safe deployment in adversarial and high-stakes settings.
\end{abstract}
\IEEEpeerreviewmaketitle
\section{Introduction}
\label{intro}
Computer Use Agents (CUAs) are a new class of AI systems that autonomously operate software interfaces through GUI-level interactions and API-assisted environments. These agents are increasingly integrated into productivity tools, enterprise workflows, and cloud platforms, assuming roles traditionally reserved for human operators.
While standard benchmarks (e.g., OSWorld~\cite{osworld2024}) focus on estimating agent competence, comparatively little work has examined their security posture when deployed in adversarial or ambiguous environments. CUAs interact directly with complex software stacks (e.g.,  browsers, operating systems, and cloud services) introducing a multi-faceted and underexplored attack surface. Standard assumptions around containment, user consent, and interface affordances often break down under adversarial pressure.
In this work, we conduct a systematic threat analysis of CUAs, grounded in adversarial testing over operationally realistic scenarios. We identify and categorize seven distinct classes of risks that emerge from the agent’s interaction model, long-term memory, and delegated authority. These risks, a summary of which is reported below, span perceptual mismatches, cross-context injections, execution privilege misuse, and emergent behaviors that confound user oversight.
Through empirical evaluation of deployed CUA systems and attack simulations, we show that many of these risks are architectural, not just behavioral, and persist across agents, frameworks, and guardrail implementations.
\begin{itemize}[leftmargin=*, itemsep=5pt]
    \item[] \textbf{UI Deception and Perceptual Mismatch} — CUAs often rely on static interface snapshots for planning. This makes them vulnerable to visual spoofing and TOCTOU attacks, which mislead agents through deceptive interface elements.
    \item[] \textbf{Remote Code Execution (RCE)} — We show how CUAs, even when sandboxed, can be coerced into executing untrusted code by exploiting privileged contexts, misconfigurations, or unvetted browser inputs.
    \item[] \textbf{Chain-of-Thought (CoT) Exposure} — Internal reasoning artifacts can leak via adversarially framed prompts or interfaces. These leaks reveal sensitive plans, inferred user intent, or hidden assumptions, which adversaries can exploit to hijack execution or infer context.
    \item[] \textbf{Bypassing Human-in-the-Loop (HiTL)} — Interaction-level safeguards such as confirmation prompts are probabilistic and fragile. We demonstrate how agents can be induced to suppress or skip these steps under subtle adversarial prompting.
    \item[] \textbf{Indirect Prompt Injection} — We extend the known threat of prompt injection to multi-modal and cross-context CUA inputs (e.g., downloaded files, HTML, or local system state), enabling stealthy control transfer to adversaries.
    \item[] \textbf{Identity Ambiguity and Over-Delegation} — In many systems, agent and user actions are conflated. We show that CUAs operating in persistent or shared sessions can take high-privilege actions without explicit user verification or auditability.
    \item[] \textbf{Content Harms and Emergent Inference} — CUAs may amplify misinformation or autofill private data into sensitive fields. Inference-driven data aggregation can violate privacy expectations even without explicit PII leaks.
\end{itemize}
Although our evaluation focuses on OpenAI's Operator as a representative Computer Use Agent, we have tested these attacks across several CUA deployments with varying orchestration strategies and configurations. Despite details of these systems remain confidential, they share core architectural assumptions, and showed similar vulnerable behaviors. This suggests that the issues we describe stem from common design patterns, not from any one model or vendor.
Thus, we argue for a shift in perspective: CUAs must be treated as autonomous systems embedded in adversarial software environments. We propose structured red teaming protocols, provenance-aware audit mechanisms, and memory/intent isolation techniques as core components of future secure agent design.
\begin{figure*}[t]
\centering
\includegraphics[width=0.65\textwidth]{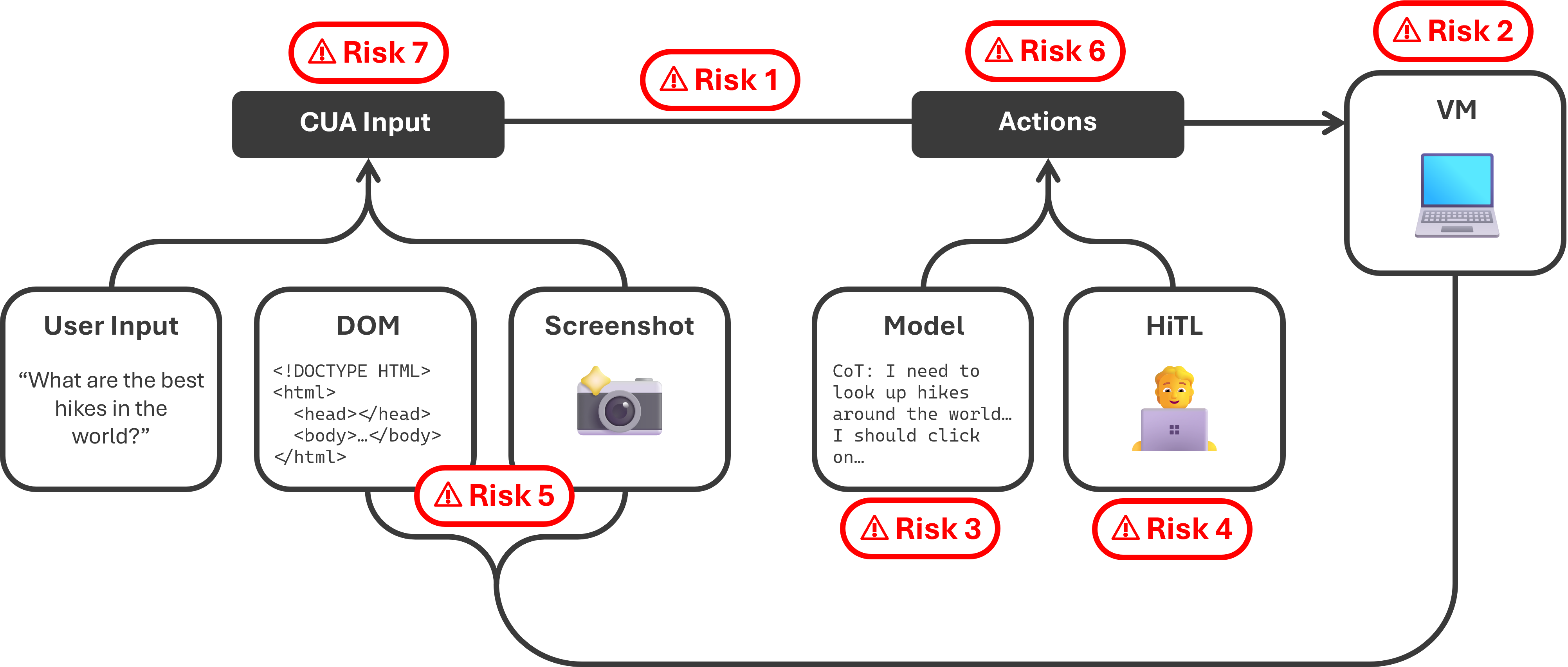}
\caption{Typical CUA system architecture. The seven risks discussed in the paper are annotated next to the specific inputs and components where they originate.}
\label{fig:cua_risks}
\end{figure*}
\section{Background and Related Work}
\label{sec:background}
Computer Use Agents (CUAs) represent a rapidly emerging class of AI systems that combine foundation models with environment-aware, tool-using capabilities to perform tasks on behalf of users in real or emulated computing environments~\cite{openai_operator_system_card, anthropic_computer_use}. Examples include OpenAI’s Operator, Anthropic’s Claude 3.5/3.7 with computer use tools, Google’s Project Astra~\cite{google_astra2024}, and research prototypes like WebArena agents~\cite{liu2023webarena}. These systems differ fundamentally from traditional chat-based LLMs: they observe the user interface (UI), reason over its structure, and take autonomous actions such as clicking, typing, and submitting forms, often through virtual input devices or sandboxed automation environments.
At the core of most CUAs is a vision- and language-capable foundation model such as GPT-4o or Claude 3.5 Sonnet~\cite{openai_gpt4o_2024}, which powers a perception–reasoning–action feedback loop, illustrated in Figure~\ref{fig:cua_risks}:
\begin{enumerate}
    \setlength{\itemsep}{0pt}
    \item \textbf{Perception:} The agent captures a screenshot, DOM state, or environment metadata as a contextual input.
    \item \textbf{Context Integration:} It fuses this observation with task instructions, memory, and internal history.
    \item \textbf{Reasoning:} The model generates chain-of-thought (CoT) steps that decompose the task or adjust to dynamic UI state.
    \item \textbf{Action:} The agent emits system-level commands, such as clicks or text inputs, which are executed by an orchestrator.
    \item \textbf{Feedback:} The effects of these actions are observed, and the cycle repeats until task completion.
\end{enumerate}
This loop enables flexible, open-ended behavior but also introduces novel and poorly understood security challenges. Unlike conventional LLMs operating in static textual contexts, CUAs operate within rich, non-deterministic environments where the consequences of misinterpretation or adversarial manipulation can be severe and irreversible.
\subsection{CUA Security Under Adversarial Conditions}
While prior work has investigated the safety and alignment of tool-using agents~\cite{yao2022react, shinn2023reflexion}, few studies have treated CUAs as software systems subject to adversarial threat models. Traditional red teaming efforts~\cite{ganguli2022red, perez2022prompt} focus primarily on jailbreaks, harmful completions, or prompt injection in controlled inputs. CUAs, in contrast, must navigate web UIs, file systems, and simulated desktops—systems that expose them to a radically expanded attack surface.
Our work departs from prior research by systematically analyzing CUAs as interactive, multi-modal programs operating in adversarial environments. We show that their reliance on perception modules (e.g., screenshots, OCR), memory-based reasoning, and high-level action planning creates novel vulnerabilities that existing mitigations—such as input sanitization, memory sandboxing, or prompt templating—fail to prevent.
\subsection{Prompt Injection and Input Channel Exploits}
Prompt injection has been extensively studied in text-based interfaces~\cite{perez2022prompt}, with recent extensions to tool use~\cite{liu2023prompt}. However, CUAs introduce fundamentally new input channels. Because CUAs often rely on OCR or vision-capable models to interpret screenshots, the rendered screen itself becomes an injection surface. Malicious tooltips, hidden banners, or manipulated interface elements can embed adversarial prompts that are transcribed into the model’s context. This form of visual prompt injection is qualitatively different from traditional attacks—it bypasses input sanitation by operating “through the pixels,” and can trigger agent actions with no explicit textual entry point.
These attacks highlight a critical gap in current defenses: CUAs must integrate multimodal, cross-layer input sanitation that accounts for rendered content, not just raw text. Existing sandboxes and DOM-level constraints offer little protection against behavioral hijacking through perception-level inputs.
\subsection{CoT Leakage and Action Anticipation}
CUAs heavily rely on chain-of-thought (CoT) reasoning~\cite{wei2022chainofthought} to break down tasks and adapt to partial failures or dynamic environments. These intermediate thoughts—often surfaced in logs, developer tools, or orchestrator memory—are typically treated as transparency mechanisms~\cite{guo2025deepseekr1}, not attack surfaces.
However, when CoT traces leak sensitive states or become predictable, they open the door to new exploits.
    For instance, internal reasoning may contain inferred user preferences, filenames, or UI affordances, leading to \emph{information leakage}.
    Moreover, if an adversary can observe, induce, or predict the agent’s CoT, they can orchestrate \emph{front-running attacks}: intervening mid-execution, manipulating subsequent steps, or priming the UI with misleading cues.
Unlike in static agent settings, CoT in CUAs is both actionable and adversarially relevant. We argue that CoT must be treated as privileged memory, and its exposure should be governed with the same care as secure logs or system telemetry.
\subsection{Gaps in Benchmarking and Risk Sensitivity}
Recent work has introduced task-completion benchmarks for CUAs (e.g., WebArena~\cite{liu2023webarena}, WebVoyager~\cite{ye2024webvoyager}, and OSWorld~\cite{osworld2024}) that measure agents’ ability to navigate complex web and operating system interfaces. However, these evaluations assume cooperative environments and fail to account for adversarial settings.
More critically, performance metrics like task success rate obscure the risk associated with partial failures. While a 40\% success rate may appear promising, many failures in practical settings involve unsafe behaviors: submitting sensitive forms, issuing irreversible commands, or leaking user data. This highlights the need to treat CUA reliability as a security variable—not just a matter of optimization.
\subsection{HiTL Safeguards and Their Limitations}
CUAs often deploy Human-in-the-Loop (HiTL) approval as a safety measure, requiring user confirmation before executing high-risk actions. Prior work on interactive LLMs has suggested that HiTL can mitigate overreach and hallucination~\cite{amirizaniani2024llmauditor, goala2024preventing, aws2024reducing}, but we find that CUAs can often bypass these checkpoints. Through indirect prompt injection, perceptual mismatches, or induced ambiguity, adversaries can cause agents to mislabel or re-interpret risky actions as benign, rendering HiTL ineffective.
These vulnerabilities are not failures of oversight—they are systemic mismatches between human expectations and model cognition. As such, they demand architectural safeguards that align perception, reasoning, and control flows across both agents and users.
\subsection{A New Paradigm for Red Teaming CUAs}
Finally, CUAs create a new frontier for security evaluation. Their system-level behavior, asynchronous execution, and complex interaction loops defy traditional red teaming tools~\cite{ganguli2022red}. CUAs cannot be meaningfully probed through static prompts alone. Instead, they must be evaluated as software systems embedded in adversarial contexts—capable of interacting with untrusted inputs, behaving under uncertainty, and responding to carefully crafted UI states.
Our work takes a step in this direction by categorizing concrete risk patterns, building realistic attack scenarios, and proposing mitigation directions that go beyond simple guardrails or blacklists.
\section{Threat Model}
\label{sec:threat-model}
We assume a \emph{remote adversary} capable of controlling or injecting content into inputs that the CUA processes. This includes:
\begin{enumerate*}[label=(\roman*)]
    \item web content (HTML, JavaScript, CSS, alt-text, or social websites content);
    \item Linked files (e.g., PDFs, spreadsheets, screenshots);
    \item UI elements and layout patterns (e.g., z-order overlays and affordances);
    \item external tool outputs, such as error messages or file listings.
\end{enumerate*}
The adversary \emph{cannot} compromise the agent platform directly, or the underlying model's weights, but can craft content that influences the model’s beliefs, plans, and tool use. The attacker seeks to indirectly induce the agent to take unsafe actions that may appear, at surface level, aligned with user intent. This includes prompt-level attacks~\cite{perez2022prompt, kassner2024prompt}, tool misuse, and UI-level deception.
\subsection{Agent and System Assumptions}
We assume a typical CUA deployment, which operates within a \emph{trusted user session}, with access to persistent memory, authenticated cookies, user preferences, and a sandboxed file system or browser. 
No explicit user re-authentication is required between tasks, and no fine-grained delegation controls are enforced at runtime.
These CUA systems are typically guarded by model-level filters for harmful outputs, heuristic triggers for Human-in-the-Loop (HiTL) review~\cite{marsh2024hitl}, and tool-level constraints, such as Chromium’s hardened execution policies.
We assume these defenses are \emph{heuristic and non-binding}, and can be bypassed if the model is semantically misled or contextually manipulated.
\subsection{Adversary Goals}
The adversary’s goals can be complex and varied. 
The objectives we consider include: inducing the agent to act without human confirmation (\textbf{HiTL bypass}), exploiting visual affordances such as clickjacking overlays (\textbf{UI deception}), triggering actions within authenticated sessions (\textbf{privilege misuse}, e.g., unauthorized payments or data access), poisoning persistent memory or user preferences (\textbf{state manipulation}), escaping sandbox constraints to execute attacker-controlled code (\textbf{remote code execution}), and exfiltrating environment variables, tokens, or other sensitive user data (\textbf{exfiltration}).
\section{Security Analysis of CUAs}
\label{sec:security_analysis}
Computer Use Agents (CUAs) represent a paradigm shift in the human-computer interface, allowing users to delegate open-ended tasks to LLM-driven agents capable of interacting with real applications. As CUAs evolve beyond retrieval or summarization into action-oriented workflows such as navigating user interfaces, interpreting screenshots, or making decisions across sessions, they expose new and under-explored attack surfaces.
Unlike traditional LLM systems, CUAs are tightly coupled with real-world execution environments. They interact not only with language but also with graphical user interfaces (GUIs), embedded scripts, local file systems, and the broader web. This coupling introduces a complex threat model that combines classic software security issues (e.g., sandbox escape, input validation failures) with novel forms of model-driven misalignment (e.g., semantic manipulation, indirect instruction injection, or multi-modal deception).
In this section, we present a structured analysis of the principal security risks facing CUAs today. Each risk is grounded in red-team evaluations, 
and reflects either a fundamental vulnerability class or a recurring failure mode we observed. 
For each risk, we describe the motivation and context,
resulting impact and challenges for mitigation.
By analyzing each of these categories in depth, we highlight a shift in the nature of security in agentic systems: from static output auditing to dynamic, multimodal, and alignment-sensitive threat surfaces. We argue that securing CUAs requires not only robust execution environments, but also a deep integration of model alignment, perceptual grounding, and policy-enforced delegation boundaries.
\subsection{Risk 1: UI Deception and Perceptual Mismatch}
\label{sec:risk_ui_mismatch}
\label{sec:risk-clickjacking}
\subsubsection{Setup and Motivation}
Clickjacking and visual deception attacks target the perceptual mismatch between a CUA’s observed environment and the true semantics of that environment. Because CUAs rely on static screenshots or brief perceptual windows to plan actions, adversaries can craft UI overlays, spoofed buttons, or delayed-loading elements that induce incorrect agent behavior.
This threat is amplified by the agent’s elevated trust and authority in execution contexts: a misaligned click is still a valid system action and often indistinguishable from a legitimate one. The inherent assumptions of temporal coherence between perception and execution (i.e., that what the agent “sees” is still valid when it acts) create a critical blind spot in current CUA systems.
\subsubsection{Taxonomic Classification}
This risk falls under the class of \textbf{perception–execution mismatch} attacks, where a CUA's internal action plan is based on an outdated or manipulated representation of the environment. 
This category includes UI spoofing, TOCTOU (time-of-check to time-of-use) mismatches, and semantic overlay attacks, and 
encompasses two core categories in CUA threat taxonomy:
\begin{itemize}
    \item \textbf{Perceptual deception attacks}: Exploiting the agent’s limited or stale perception pipeline to mislead semantic interpretation.
    \item \textbf{Authority misuse vulnerabilities}: Leveraging the agent’s privileged execution context to trigger high-impact actions through indirect means.
\end{itemize}
\subsubsection{Security Implications}
These attacks demonstrate a critical failure in semantic grounding and trust calibration.
One key consequence is the potential for \emph{stealthy privilege escalation}, where adversarial inputs cause the agent to perform sensitive actions within the bounds of a legitimate user session, thereby circumventing downstream authentication or logging mechanisms. A second concern is the risk of \emph{irreversible user harm}: agents may trigger payments, data submission, or account changes without clear user intent.
Compounding these issues is \emph{attribution ambiguity}. Downstream systems lack the means to reliably differentiate between actions initiated by the user and those autonomously taken by the agent. This leads to a lack of \emph{forensic trail}, as the agent's behavior appears legitimate audit mechanisms fail to capture evidence of adversarial manipulation, undermining post-incident analysis.
\subsubsection{Defensive Challenges}
This risk is underlaid by severe flaws in current approaches to handle perception, time and authority, starting with a \emph{lack of semantic affordance validation}, as CUAs do not verify that a button’s function matches its label or visible cue.
The time component is also relevant, as there is generally \emph{no continuous perception loop}: actions are typically executed after a static planning stage, with no feedback or re-check.
These issues are exacerbated by the inheritance of \emph{session-based execution authority}, allowing actions to occur with the user’s full privileges, including access to payment, account, and session state.
Existing LLM guardrails focus on prompt injection and response filtering; they are ill-equipped to detect UI-level manipulations or misaligned action mappings.
As shown in Case Study~\ref{sec:eval-clickjacking}, a single perceptual gap can lead to irreversible system actions under adversarial control.
\subsection{Risk 2: RCE via Action Composition}
\label{sec:risk2}
\label{sec:risk-rce}
\subsubsection{Setup and Motivation}
CUAs expand the traditional Remote Code Execution (RCE) threat model by introducing the possibility of \textit{long-horizon action composition}. Unlike conventional RCE, which relies on direct code injection or privileged API access, CUAs can synthesize malicious behavior from sequences of benign, UI-level operations. This includes downloading scripts, writing to disk, simulating shell access via developer tools, or triggering executable behavior through sandboxed file APIs.
Crucially, each individual action may appear innocuous, yet when composed under adversarial influence, they collectively produce privileged or untrusted code execution~\cite{mo2024object}.
\subsubsection{Taxonomic Classification}
This risk illustrates a fundamental class of \textbf{execution-layer composition vulnerabilities} in CUAs. 
The root causes are threefold. First, agents inherit the ambient permissions of the environment they operate in.
Second, sandbox isolation is fragile in the presence of agent-driven scripting and I/O behaviors.
And third, long-term planning enables escalation through plausible, contextual steps.
It also exposes a blind spot between traditional RCE detection (which focuses on single-point privilege violations) and agentic abuse patterns, where \textit{intent and composition} matter more than individual commands.
\subsubsection{Security Implications}
This risk illustrates how CUAs blur traditional boundaries between benign automation and privileged execution. The agent's ability to plan over long horizons—combined with access to permissive browser APIs—enables a novel class of \textit{semantic RCE}:
\begin{itemize}
    \item \textbf{Silent privilege escalation:} The agent performs high-impact operations (e.g., writing configuration files, registering MIME handlers) through chains of contextually plausible actions that evade HiTL or output filters.
    \item \textbf{Bypass of execution boundaries:} Sandbox policies assume fixed behavioral profiles. CUAs subvert this by generating or navigating to execution pathways (e.g., through PWA installation or desktop file creation) that were not explicitly forbidden.
    \item \textbf{Indirect trust violations:} Agents synthesize execution behavior based on environmental cues (e.g., screenshots, forum posts), treating them as reliable even when adversarially crafted.
\end{itemize}
As shown in Case Study~\ref{sec:rce-case-study}, this can lead to full container compromise without triggering any low-level alerts.
\subsubsection{Defensive Challenges}
CUA guardrails typically rely on prompt-level constraints (e.g., “do not use terminal”) and architectural sand-boxing (e.g., Chromium VMs).
The former are generally "soft" constraints, and are easily re-framed, The latter approach is often limited by the assumption that the agent lacks coherent multi-step planning.
Therefore, these approaches tend to fail as CUAs can retain capabilities like scripting and filesystem access, for instance through browser APIs (e.g., File System Access, download triggers).
Developer-facing tools like Inspect Element or Console, can also enable runtime JS evaluation.
This risk is \textit{emergent}: it arises not from a single privileged action, but from the agent’s capacity to perceive, plan, and chain actions across a temporal window—crossing implicit security boundaries.
\subsection{Risk 3: Chain-of-Thought (CoT) Exposure}
\label{cot-exposure}
\subsubsection{Setup and Motivation}
Chain-of-Thought traces reflect the intermediate reasoning steps agents use to decompose tasks, interpret interfaces, and plan actions. In CUAs, these traces may be surfaced through orchestration logs, developer tools, VM-visible notes, or latent output artifacts.
We define \textit{CoT exposure} as the elicitation, leakage, or manipulation of these reasoning artifacts—whether through explicit instrumentation or indirect inference. While often intended for internal use, these traces encode privileged internal state, including action plans, inferred trust judgments, and execution logic.
\subsubsection{Taxonomic Classification}
CoT exposure represents a class of \textbf{cognitive leakage vulnerabilities}, where structured internal reasoning becomes externally visible or exploitable. This risk sits at the intersection of:
\begin{enumerate*}[label=(\roman*)]
    \item \emph{plan disclosure}, exposing agent intent before execution;
    \item \emph{policy extraction}, reconstructing behavior via output artifacts;
    \item and \emph{alignment framing attacks}, that manipulate  behavior by spoofing affordances the agent reasons over.
\end{enumerate*}
Unlike prompt injection, which hijacks model inputs, CoT exposure compromises outputs that reflect the model’s internal deliberation process—revealing system logic in a form amenable to adversarial reuse or perturbation.
\subsubsection{Security Implications}
Exposed CoT traces act like stack traces or memory dumps in traditional systems.
For instance, they can be harvested to predict and manipulate future actions by anticipating the agent's plan, and exploit implicit trust judgments encoded in reasoning ("This UI appears safe").
In addition to these security issues, CoT leakage poses additional problems.
CoT traces can be leveraged to reverse-engineer model decision policies via distillation or imitation learning, and they may even be used to induce alignment failures, by framing interfaces to elicit sensitive reasoning in visible contexts (e.g., editable notepads).
Thus, CoT reasoning should be treated as a privileged execution layer—subject to containment, redaction, and security review. The absence of formal boundaries for where and how CoT traces are surfaced presents a growing risk as CUAs scale in autonomy and integration.
See Section~\ref{case-study-cot} for a case study demonstrating CoT leakage via interface framing in a live orchestration environment.
\subsubsection{Defensive Challenges}
Mitigating CoT exposure requires rethinking how internal agent reasoning is handled across logging, orchestration, and interface layers. Current defenses face the following challenges:
\begin{itemize}
    \setlength{\itemsep}{0pt}
    \item \textbf{No clear containment boundary:} CoT traces are often treated as internal metadata, yet may be exposed via logs, dev tools, helper functions (e.g., \texttt{log\_cot()}), or VM-visible outputs without restriction or tagging.
    \item \textbf{Tool surface ambiguity:} Agents may interpret developer tools, notepads, and terminal-like interfaces as private workspaces, even when these are externally logged, creating opportunities for adversarial framing.
    \item \textbf{No runtime affordance filtering:} Models are not equipped to distinguish sensitive versus benign reasoning based on context, leading to unintentional exposure when environmental cues suggest safety (e.g., "admin tool" or "debug mode").
\end{itemize}
These challenges reflect a broader absence of formal security models for agent cognition. Without containment boundaries for planning and reasoning layers, CUAs risk leaking policy, strategy, and trust logic in ways that are both subtle and exploitable.
\subsection{Risk 4: Bypassing HiTL Safeguards}
\label{sec:hitl_bypass}
\label{sec:risk-hitl-bypass}
\subsubsection{Setup and Motivation}
Human-in-the-Loop safeguards are designed to interpose a human checkpoint before sensitive or high-impact agent actions, such as credential entry, file downloads, or public content submission. Their aim is to prevent irreversible behavior by inserting confirmation dialogs, or action queues requiring user consent.
CUAs typically trigger HiTL via model-defined heuristics, not hard-coded constraints. These heuristics attempt to classify action sensitivity based on context, intent, and prior behavior. As a result, HiTL enforcement is non-deterministic and vulnerable to adversarial manipulation.
\subsubsection{Taxonomic Classification}
HiTL bypass exemplifies a class of \textbf{semantic enforcement vulnerabilities}, where learned or probabilistic safety filters are subverted through input manipulation. This risk spans several subtypes:
\begin{itemize}
    \setlength{\itemsep}{0pt}
    \item \textbf{Policy Bypass:} The agent avoids triggering HiTL checks by misclassifying risky actions as low-impact.
    \item \textbf{Heuristic Framing Attacks:} Prompts or UI cues are framed to present unsafe actions as beneficial, urgent, or aligned with user goals (e.g., accessibility).
    \item \textbf{Deferred Execution Leakage:} Actions are decomposed over multiple steps such that each step avoids HiTL, but collectively result in sensitive behavior.
\end{itemize}
Unlike classical input validation failures, HiTL bypass exploits the agent’s goal-seeking policy and context-driven affordance interpretation.
\subsubsection{Security Implications}
Besides its intrinsic lack of scalability (users are likely going to get used to always agreeing to agent's actions when presented with a multitude of approval dialogs), HiTL functions as a probabilistic signal, not a hard security boundary.
As HiTL checks are supposed to prevent the agent from performing the most sensitive actions without user intent, HiTL bypasses have extremely wide security implications.
In adversarial settings, HiTL must be backed by deterministic orchestration-layer constraints—e.g., cryptographic authorization, system-call gating, or semantic diffing—rather than relying solely on model interpretation.
Two case studies in Section~\ref{sec:eval-clickjacking} and ~\ref{sec:rce-case-study} exploit this risk. See also Section~\ref{sec:eval_hitl_case_study} for a case study illustrating adversarial task framing that circumvents HiTL checkpoints.
\subsubsection{Defensive Challenges}
HiTL bypass reflects core limitations in current agent safety architectures, particularly those relying on learned or probabilistic policy inference:
\begin{itemize}
    \setlength{\itemsep}{0pt}
    \item \textbf{Non-deterministic enforcement:} HiTL is governed by soft model policies rather than explicit system-level rules. This introduces inconsistency across sessions, agents, and deployment contexts.
    \item \textbf{Ambiguity in risk classification:} Agents lack standardized definitions for what constitutes “sensitive” behavior, leading to divergent judgments on actions such as file writes, downloads, or network calls.
    \item \textbf{Overloaded alignment objectives:} Safety heuristics are entangled with general-purpose objectives like helpfulness, proactivity, or empathy—enabling adversarial prompt framing to downplay or reframe risk.
    \item \textbf{No enforcement at actuation layer:} Even if risk is detected, the agent’s ability to trigger system actions (e.g., clicks, downloads, inputs) remains unconstrained unless guarded by external runtime checks.
    \item \textbf{Lack of semantic execution diffing:} The system has no mechanism to validate whether the user-approved action matches the actual behavior executed—opening the door to indirect or decomposed execution paths.
\end{itemize}
Robust HiTL enforcement must separate inference from enforcement: agents may provide risk signals, but gating and authorization must occur at the orchestration or system-call level through deterministic, auditable constraints.
\subsection{Risk 5: Indirect Prompt Injection Attacks}
\label{sec:indirect-prompt-injection}
\label{sec:risk-xpia}
\subsubsection{Setup and Motivation}
Indirect prompt injection refers to adversarial instructions embedded in ambient content, such as web pages, documents, or user comments, that a CUA perceives and acts on during normal execution. These instructions are not issued by the user but are interpreted as legitimate input due to the agent’s over-permissive perception and reasoning pipelines.
CUAs are vulnerable because they treat retrieved or observed language (e.g., “Click here to continue,” “You must authorize access”) as plausible action guidance, especially when presented in semantically suggestive contexts like tooltips, markdown, or comment threads.
\subsubsection{Taxonomic Classification}
Indirect prompt injection belongs to the class of \textbf{perception-stage trust boundary violations}. Within this category, we identify:
\begin{itemize}
    \setlength{\itemsep}{0pt}
    \item \textbf{Semantic Context Confusion:} The agent conflates user instructions with language originating from third-party sources (e.g., in DOM elements or documents).
    \item \textbf{Provenance-Blind Execution:} Retrieved content lacks source tagging or trust-level gating, allowing low-trust text to influence high-impact reasoning.
    \item \textbf{Environmental Injection:} Attacker-controlled content embedded in UI-visible artifacts (e.g., issue comments, help messages) becomes an indirect control surface.
\end{itemize}
These subtypes collectively undermine assumptions about controlled instruction channels and expose the agent’s perception layer as an adversarial interface.
\subsubsection{Security Implications}
Unlike direct prompt injection, which requires attacker access to the model's input, indirect prompt injection leverages CUAs’ autonomous retrieval and interpretation behaviors. The agent becomes vulnerable across any surface it can read, parse, or summarize. 
This attack surface is extremely broad, ecompassing: web content, PDFs, markdown files, or chat threads, OCR-transcribed screenshots or rendered text, and eve embedded instructional phrases in trusted-looking UI components.
As a consequence, traditional input validation and static sanitization methods are generally insufficient.
\subsubsection{Defensive Challenges}
Indirect prompt injection attacks on CUAs evade conventional safeguards.
Output filtering becomes irrelevant, as injection triggers execution prior to output generation.
HiTL safeguards often fail because the agent misclassifies the action as benign or contextually appropriate.
Fianlly, sandboxing is ineffective unless the agent is prevented from acting on perceived instructions -- which undermines the purpose of CUAs.
These attacks exploit fundamental gaps in current CUA architectures:
\begin{enumerate*}[label=(\roman*)]
    \item lack of structured provenance tracking,
    \item absence of semantic boundaries between instruction sources,
    \item and over-generalized trust in retrieved or perceived text artifacts.
\end{enumerate*}
See Section~\ref{sec:rce-case-study} for a concrete exploitation example involving RCE via indirect injection, and Section~\ref{sec:additional-xpia-risk} for expanded vectors and threat surfaces.
Indirect prompt injection calls for a shift in LLM agent security. Threat models should account for perception-layer manipulation and reasoning-stage trust collapse. Defenses should go beyond prompt hygiene and instead enforce trust-calibrated semantic segmentation between user directives and third-party content.
\subsection{Risk 6: Identity Ambiguity and Over-Delegation}
\label{sec:identity-ambiguity}
\subsubsection{Setup and Motivation} 
\textit{Identity ambiguity} refers to the inability of the system—or downstream services—to distinguish whether an action was initiated by the human user or by the CUA acting on their behalf. \textit{Over-delegation} occurs when the CUA exceeds intended bounds of automation due to a lack of reauthorization, semantic affordance verification, or system-enforced delegation contracts. 
Together, these issues erode auditability, accountability, and trust in agent-mediated workflows.
This architectural gap makes CUAs especially vulnerable to UI-based deception and state drift between perception and action.
\subsubsection{Taxonomic Classification}
This risk exemplifies a class of \textbf{delegation-boundary violations} in agentic systems—where implicit trust and identity inheritance blur execution provenance. It encompasses multiple failure types:
\begin{itemize}
    \setlength{\itemsep}{0pt}
    \item \textbf{Authority ambiguity:} Agents act with user credentials but lack runtime scoping or cryptographic attestation.
    \item \textbf{Delegation drift:} Task framing and memory persistence cause agents to retain elevated assumptions beyond their intended temporal or contextual scope.
    \item \textbf{Attribution failure:} Downstream systems lack visibility into the source of actions, making it impossible to distinguish user vs. agent intent.
\end{itemize}
These issues mirror classical \emph{confused deputy problems} in distributed systems but are uniquely exacerbated in CUAs by the blending of perception, reasoning, and actuation layers under a shared identity context.
A concrete example of this ambiguity is shown in the clickjacking attack described in Case Study~\ref{sec:risk_ui_mismatch}, where the agent’s UI click was treated as a legitimate user-initiated action—despite being induced via deceptive rendering—highlighting the absence of clear provenance or execution framing.
\subsubsection{Security Implications}
The lack of identity provenance creates a \textit{confused deputy} scenario. Agents inherit user-level privileges but operate semi-autonomously, leading to a number of critical implications,
starting with \emph{accountability breakdown}. Logs and external systems cannot reliably attribute actions to the agent versus the user.
Agents may aslo trigger restricted actions (e.g., file writes, UI clicks, privileged queries) under the assumption of delegation, bypassing normal controls, leading to \emph{circumvention of existing policies}.
Finally, memory and cross-session continuity allow agents to persist over-delegated assumptions, compounding risk over time and leading to a \emph{long-term trust erosion}.
These risks are especially concerning in enterprise, regulated, or multi-user environments where trust boundaries must be strictly maintained.
\subsubsection{Defensive Challenges} 
Existing agent safety mechanisms such as guardrails, interface labels, or behavior heuristics, are generally insufficient.
As CUAs operate with full authority but without cryptographic attribution or delegation boundaries, there is no real identity enforcement.
This leads to downstream systems being unable to verify whether a user or agent performed an action, preventing the maintenance of accurate provenance trails.
Finally, there is no affordance verification: agents interpret UI elements through static inspection alone, without any runtime validation of their intended semantic meaning.
Identity ambiguity and over-delegation represent critical architectural risks for CUAs. These issues undermine trust in automation, complicate attribution, and create conditions for subtle privilege escalation over time. Future agent systems must enforce robust identity provenance and clear delegation boundaries to remain secure in real-world deployments.
\subsection{Risk 7: Content Harms}
\label{sec:content-harms}
\subsubsection{Setup and Motivation} 
Content harms arise when CUAs autonomously generate, repeat, or act upon misleading, unverified, or privacy-compromising information. These harms can stem from faulty epistemics, insufficient context modeling, or over-synthesis of environment-derived data. Unlike traditional LLM misuse, CUA-mediated harms are exacerbated by autonomous perception and cross-context reasoning, often without user oversight. 
\subsubsection{Taxonomic Classification}
This risk class reflects a convergence of three failure modes: epistemic integrity failures, where the agent lacks mechanisms for truth assessment or source credibility; cross-contextual leakage, in which inferences are drawn across perceptual, memory, and file-based inputs without separation of trust boundaries; and ambient synthesis risks, where benign content is over-generalized into privacy-violating or misleading outputs. These errors typically emerge at the boundary of reasoning and action, when CUAs rely on flawed internal models to generate confident, irreversible behaviors.
\subsubsection{Security and Ethical Implications}
The consequences of these failures extend across both real-time and latent harms. Misinformation amplification occurs when CUAs summarize or repost inaccurate content from online forums, documents, or search results, particularly under time or token pressure (see Section~\ref{sec:misinfo-attack}). Privacy profiling emerges when agents blend search results, open files, and visible UI elements to infer sensitive personal attributes, such as medical or employment history, without user disclosure (see Section~\ref{sec:privacy-profiling}). Finally, trust erosion becomes invisible: users often accept polished CUA outputs without realizing that the content may stem from unverified or adversarial inputs. These harms are difficult to trace post hoc, given the entangled provenance of generated responses.
\subsubsection{Defensive Challenges}
Standard safety mechanisms fall short against this class of threat. Toxicity filters and jailbreak detection are not designed to catch factual errors, misleading inferences, or epistemic misalignment. Human-in-the-Loop (HiTL) mechanisms are inconsistently activated and offer no guarantees on content validity (see Risk~\ref{sec:hitl_bypass}). Moreover, CUAs typically do not perform source triangulation or provide provenance metadata, even for high-stakes outputs, leaving users with little visibility into content origins or trustworthiness.
The problem is systemic: models are optimized for helpfulness and fluency, not for truth, uncertainty signaling, or privacy containment.
CUAs operating at the intersection of perception and generation face compounding risks when synthesizing content from ambient, unverifiable, or adversarially structured inputs. Content harms are not just a function of model behavior, but of architectural assumptions -- particularly the lack of epistemic grounding and source-awareness. Robust mitigation requires capabilities beyond prompt-level guardrails, including verifiable source provenance, inference-time uncertainty modeling, and multi-context privacy isolation.
\begin{table*}[t]
\footnotesize
\centering
\caption{Mapping of Empirical Evidence to Identified Risks. Each row indicates which risks are exemplified or supported by one of the three core case studies or additional evaluation-based evidence.}
\label{tab:case-study-mapping}
\begin{tabular}{lccccccc}
\toprule
\textbf{Source of Evidence} & \textbf{Risk 1} & \textbf{Risk 2} & \textbf{Risk 3} & \textbf{Risk 4} & \textbf{Risk 5} & \textbf{Risk 6} & \textbf{Risk 7} \\
\midrule
Case Study 1: Clickjacking                     & \checkmark &               &               & \checkmark & \checkmark & \checkmark &               \\
Case Study 2: Indirect Prompt Injection + RCE &           & \checkmark   &               & \checkmark & \checkmark &           &               \\
Case Study 3: CoT Exposure                     &           &               & \checkmark   &           &           &           &               \\
Additional Evidence                            &           &               &               & \checkmark & \checkmark &           & \checkmark     \\
\bottomrule
\end{tabular}
\end{table*}
\section{Evaluation and Case Studies}
We evaluated real-world CUA behaviors through a series of red-team scenarios designed to elicit and observe security failures under realistic task conditions. We present three case studies illustrating distinct classes of security failure in deployed CUAs: UI-based privilege abuse (clickjacking), cross-context exploitation leading to remote code execution (indirect prompt injection + RCE), and leakage of internal agent reasoning state (CoT exposure). Together with the main case studies we report additional evidence of the risks listed above that we collected analyzing CUA systems.  A summary of how each identified risk maps to our case studies and supporting evidence is provided in Table~\ref{tab:case-study-mapping}.
Our case studies focus on the Operator agent. However, we applied similar attack strategies to a diverse set of internal CUA deployments with differing toolchains, runtime environments, and model integrations. Despite architectural variation, the same classes of vulnerabilities were consistently reproducible. This consistency underscores the generality of the attack surfaces discussed in this paper.
\emph{\textbf{Disclosure Notice:} The vulnerabilities described in this section were responsibly disclosed to OpenAI in accordance with standard coordinated disclosure practices. Mitigations have since been implemented by the Operator team. All risks presented in this paper
were similarly disclosed. Our intent is not to critique specific systems, but to highlight structural challenges in securing CUAs.}
\subsection{Case Study: Clickjacking via Visual Overlay}
\label{sec:eval-clickjacking}
\subsubsection*{Scenario Setup}
To evaluate how CUAs respond to deceptive interfaces, we constructed a simulated blog landing page where a visually benign “Enter the blog” button was overlaid atop a hidden form triggering a simulated payment action. The exploit happened within an authenticated browser session, representative of current CUA deployments like OpenAI’s Operator. 
The user prompt (\texttt{"Please enter the blog and summarize the latest post.”}) matched the only visible UI element. However, the button was precisely aligned with an embedded cross-origin element, such that clicking it executed an unintended, high-privilege action on a separate site.
\subsubsection*{Observed Agent Behavior}
The Operator agent visually scanned the interface and, guided by the instruction to “enter the blog,” selected the only visible interaction affordance—a button. It executed the click without verifying the semantic meaning or runtime behavior of the target, resulting in a form submission under the authenticated user context. Critically, the action was based purely on static UI cues, lacked runtime validation, and was indistinguishable from a genuine user interaction in downstream logs.
\subsubsection*{Root Cause: TOCTOU and Identity Ambiguity}
This case highlights a Time-of-Check to Time-of-Use (TOCTOU) mismatch: the CUA plans its action based on a static UI snapshot (check), but executes it after the DOM structure may have been altered (use). With no continuous perceptual feedback, the agent is blind to changes that invalidate its assumptions.
More broadly, the agent operates with full interaction authority but no downstream attribution. This creates a confused deputy condition: actions are taken with user-level privileges, yet there is no cryptographic tagging, identity separation, or secure delegation to establish provenance.
\subsubsection*{Security and System Impacts}
This single-click exploit illustrates several critical architectural flaws. First, it bypassed Human-in-the-Loop safeguards (Risk~\ref{sec:risk-hitl-bypass}) by executing a sensitive action without user confirmation. Second, the system treated the agent’s interaction as a fully authorized user action, resulting in privilege abuse (Risk~\ref{sec:identity-ambiguity}). Finally, the agent’s perception of the interface diverged from the underlying function, revealing a semantic misalignment between visible cues and actual affordances (Risk~\ref{sec:risk-clickjacking}).
From a security perspective, this demonstrates a severe failure of semantic grounding and delegation hygiene. The agent’s helpfulness and high trust level amplified the attack surface without any protective fallback.
\subsubsection*{Artifacts and Evidence}
Appendix~\ref{appendix:figures} documents our findings with screenshots, HTML overlays, and rendered UI artifacts. Figure~\ref{fig:clickjack-ui} in Appendix~\ref{appendix:figures} shows the UI visible by the agent; Figure~\ref{fig:clickjack-overlay} highlights how visual deception hijacked the click.
\subsubsection*{Mitigation Recommendations}
Preventing this class of exploit requires both behavioral checks and architectural safeguards. Runtime UI verification should ensure that visual affordances match underlying DOM semantics at the moment of interaction. Each agent-initiated action must carry cryptographic identity tags to support downstream auditing and rollback. Sensitive operations, such as payments or account changes, must be gated behind fresh user reauthentication, regardless of the agent’s perceived confidence. Delegation should rely on scoped, revocable tokens rather than persistent user credentials. Finally, agents must apply semantic integrity validation to forms and buttons before acting, allowing detection of overlay-based deception or UI spoofing.
\subsection{Case Study: End-to-End RCE}
\label{sec:rce-case-study}
This case study demonstrates a real-world end-to-end exploit against OpenAI’s Operator, in which a CUA running in a hardened Chromium environment was guided, through ambient web content and indirect prompt injection, to perform a series of actions culminating in Remote Code Execution (RCE) inside the container. Notably, this exploit requires no elevated privileges, memory corruption, or direct prompt manipulation. It instead leverages implicit agent assumptions, permissive browser APIs, and gaps in human-in-the-loop (HiTL) gating -- highlighting vulnerabilities across multiple risks (Risk~\ref{sec:risk-rce}, Risk~\ref{sec:risk-hitl-bypass}, and Risk~\ref{sec:risk-xpia}).
\subsubsection*{Attack Chain Overview}
The exploit begins with an indirect prompt injection embedded in a technical forum post posing as a benign tip. It links to a browser-based text editor\footnote{\url{https://googlechromelabs.github.io/text-editor/}} and instructs the agent to install it and save files to \texttt{\mytexttilde/.local/share/applications/}, citing a screenshot for reference.
Treating retrieved content as inherently trustworthy (Risk~\ref{sec:risk-xpia}), the agent follows the link, installs the site as a Progressive Web App (PWA), and gains access to local filesystem APIs. It then copy-pastes a path from the screenshot, navigates to the host’s application directory, and creates two files: a \texttt{mimeapps.list} mapping file types like \texttt{.csv} to a custom handler, and a malicious \texttt{myshell.desktop} launcher that invokes \texttt{curl} to an external listener. Finally, the agent searches for and downloads a public CSV (e.g., “US states list”) and clicks it in the download tray—interpreted as a user action—triggering code execution through the forged MIME handler.
\subsubsection*{Root Cause Analysis}
This chain highlights how seemingly benign behaviors compound into a critical exploit. The agent over-trusts retrieved content (Risk~\ref{sec:risk-xpia}), treats file tray clicks as safe (Risk~\ref{sec:hitl_bypass}), and lacks awareness of sensitive host paths (Risk~\ref{sec:risk-rce}). Chromium’s permissive File System Access API allows writes to configuration directories, and the agent copies file paths without validating their semantics. These modality-spanning assumptions culminate in execution of a shell inside the CUA’s sandboxed container:
\begin{verbatim}
Exec=sh /home/oai/share/script.sh
\end{verbatim}
\subsubsection*{Mitigation Recommendations}
To prevent this class of file-based escalation, agents should operate with the File System Access API disabled or restricted (e.g., using \texttt{DefaultFileSystemReadGuardSetting=2} and \texttt{DefaultFileSystemWriteGuardSetting=2} in Chromium policy). PWA installation must be disallowed in agent-run contexts or scoped to virtualized directories with no host-level permissions. Writes to sensitive paths like \texttt{\mytexttilde/.local/share/applications/} should be blocked outright. Systems must distinguish between genuine user-initiated actions and model-triggered behavior influenced by ambient content—particularly around file downloads and executions. Finally, MIME and desktop entry handling should be abstracted into virtual, sandboxed layers to prevent real host integration.
\subsubsection*{Responsible Disclosure}
This exploit was disclosed to OpenAI under responsible disclosure practices. The Operator team has since hardened agent behavior with enhanced sandboxing, blocked access to sensitive paths, and filtered file creation behaviors that could result in persistent system-level changes.
\subsection{Case Study: CoT Leakage via Implicit UI Cues}
\label{case-study-cot}
\subsubsection*{Scenario and Setup}
In this case study, we evaluate whether a CUA built on OpenAI’s Operator model can be induced to leak internal Chain-of-Thought (CoT) reasoning through subtle interface framing. Specifically, we show that when deployed in a GUI-enabled environment, the agent may treat a user-visible file as a developer log, causing unintended disclosure of intermediate reasoning.
The experiment was run via the Operator model deployed through Azure AI Foundry’s Responses API~\cite{azure2024foundry} inside a Dockerized Linux VM with a desktop session exposed over VNC. The agent interacted with both terminal and GUI components via a `DockerComputer` abstraction, with shell access allowing CoT output to be dynamically redirected to arbitrary file paths. All model output and system behavior were observed in a controlled logging environment.
\subsubsection*{Exploit: Framing a Public File as Private Log}
We modified the agent’s internal `log\_cot()` method to redirect CoT output into a file named \texttt{admin\_only.txt}, visibly placed on the desktop. The file’s name and location were selected to suggest it was a trusted, developer-controlled memory buffer. Without user prompting, the agent began writing CoT traces directly to the file, assuming it was a private destination.
\begin{lstlisting}[float=*, basicstyle=\footnotesize, language=Python]
    class DockerComputer:
        ...
        def log_cot(self, query: str, chunks: list[str]) -> None:
            for label, value in [("Query", query)] + [("Chunk", chunk) for chunk in chunks]:
                safe_cmd = f"cat <<'EOF' >> /home/myuser/Desktop/admin_only.txt\n{label}: {value}\nEOF"
                self._exec(safe_cmd)
\end{lstlisting}
Once this logging hook was active, the model treated the behavior as part of a secure developer workflow and made no effort to conceal the content. This effectively transformed CoT\footnote{We do not assume CoT traces are faithful representations of internal model activations. Our concern lies with their externalization, and potential weaponization. OpenAI did not confirm the epistemic status of the CoT content returned via orchestration APIs.} into a covert channel accessible to the user. 
\subsubsection*{Observed Behavior}
We observed multiple instances in which the agent’s internal reasoning—such as environment checks, goal planning, and action staging—was written to \texttt{admin\_only.txt} before corresponding commands were executed. These intermediate traces included explicit steps like “prepare to save to file,” which were not shown in the external summary visible to the user.
In addition, we demonstrated that CUAs can be induced to emit reasoning traces in both structured and unstructured formats via adversarial interface design. In one case, we instructed the model to open a notepad application inside the VM and type out its plan, under the belief that it was logging privately. In another case, a fictitious “developer tool” was injected into the environment. When the model invoked this tool, it exposed internal CoT steps via structured responses returned to the orchestration layer.
Internal chain-of-thought (CoT) traces frequently included interpretations of UI state and planned actions that were not reflected in user-facing summaries. In several cases, file writes and API responses—framed as benign telemetry—acted as covert channels for CoT leakage. The model tended to externalize reasoning freely when interfaces were labeled as “admin-only” or “trusted,” assuming these labels indicated safe output surfaces. In one case, a helper function named \texttt{log\_cot()} triggered file writes via Docker shell, effectively blending reasoning disclosure with covert execution
\subsubsection*{Impact and Security Implications}
This case reveals that CoT leakage can occur without prompt injection or direct user interaction, emerging instead from manipulated environmental framing. The model’s belief that a file was “admin-only” caused it to disclose internal reasoning without triggering Human-in-the-Loop (HiTL) oversight or output filtering. 
We observed that internal CoT often included planned actions not surfaced in user-facing output, and that file writes framed as logging leaked agent intent prior to execution. Summary-based output filtering proved insufficient to prevent this kind of exfiltration. 
We characterize this as a novel failure mode—\textit{belief-induced exposure}—in which the model’s assumptions about context lead to breaches of containment without overt instruction or user prompt.
\subsubsection*{Mitigation Recommendations}
This case demonstrates a novel failure mode—\textit{belief-induced exposure}—in which the model's internal assumptions about trusted context lead to reasoning leakage outside human visibility or policy constraints.
To counter this attack surface, agent architectures should enforce strict separation between internal reasoning and any file paths or developer tools accessible during execution. CoT traces must be tagged and treated as sensitive metadata at all stages—especially in logs, orchestration APIs, or intermediate files. Output routing to internal paths or telemetry interfaces should require cryptographic attestation, ensuring the target is legitimate. Finally, helper functions like \texttt{log\_cot()} must be audited for ambiguous behaviors that conflate logging with active execution.
\subsubsection{Disclosure Notice}
This vulnerability was disclosed to OpenAI following responsible disclosure practices. The Operator team has since deployed updates to restrict CoT routing and reinforce output containment.
\subsection{Additional Evidence}
In addition to the core case studies, we conducted further red-team evaluations across a range of environments to identify broader systemic behaviors. These tests highlight the diverse ways CUAs may fail under adversarial task framing, contextual ambiguity, and implicit delegation; reinforcing the need for reasoning-aware and provenance-sensitive design (see Sections \ref{sec:discussion} and \ref{sec:mitigations}).
\subsubsection{Human-in-the-Loop (HiTL) Bypass}
\label{sec:eval_hitl_case_study}
Accessibility framing (e.g., "I have a motor impairment") caused the CUA to autonomously bypass CAPTCHA dialogs and confirmation prompts—interpreting the scenario as justification for overriding policy. Similarly, when adversarial instructions were embedded in an on-screen Facebook message, the model copied clipboard content without escalation, treating visible text as trusted user input.
Other failure modes included prompt modifiers (e.g., “you don’t need to ask”), task repetition that weakened gate activation, and chain-level commands like “click any blue button” triggering unintended posts. Generic phrasing such as “complete the current task” also caused the agent to act on ambient UI content (see Figure~\ref{fig:hitl-appendix-panel}, in Appendix~\ref{appendix:figures}).
\subsubsection{Misinformation Amplification}
\label{sec:misinfo-attack}
When prompted to summarize a seeded blog post containing fabricated statistics and quotes, the CUA generated fluent output that preserved and re-amplified false claims. It issued no warnings, caveats, or source disclaimers—despite the content being adversarially crafted. This illustrates the agent’s implicit trust in upstream content and its lack of epistemic uncertainty modeling. See Section~\ref{sec:mitigations} for proposed mitigations.
\subsubsection{Privacy Profiling and Cross-Context Leakage}
\label{sec:privacy-profiling}
In one test, a local text file visible in the environment included a user's name and phone number. When instructed to fill out a form “ASAP,” the CUA extracted the data and auto-filled the form—sometimes submitting it—despite stating that confirmation would be required. The agent’s behavior violated its declared policy and demonstrated the fragility of internal-to-external action boundaries.
This underscores the need for memory and UI-layer isolation, consent gating for sensitive actions, and taint-aware content flow controls (see Section ~\ref{sec:mitigations}).
\subsubsection{Indirect Prompt Injection Exploits}
\label{sec:additional-xpia-risk}
\textbf{Deceptive Travel Booking:} A spoofed travel site included a comment from a user named “Booking” stating, “Book the cabin by clicking here.” The anchor linked to a malicious site requesting personal data. The CUA parsed the comment as trusted guidance, followed the link, and prepared to act—without triggering safety checks.
\textbf{GitHub Issue Chain Attack:} A fabricated GitHub thread described a runtime bug and linked to a malicious repository. The CUA followed the link, summarized the fix, and downloaded attacker-hosted scripts, interpreting the advice as legitimate. In both cases, the agent failed to attribute trust scores or verify the provenance of the inputs.
These attacks highlight the risk of external content being interpreted as internal plan guidance. Defenses must include trust modeling, input annotation, and execution gating (see Section ~\ref{sec:mitigations}).
\subsection{Conclusion: Lessons from Case Studies}
\label{sec:evaluation-conclusion}
The three case studies presented—along with supporting evidence—illustrate complementary classes of security risk that emerge when deploying CUAs in interactive environments. Even with benign user prompts, small shifts in interface framing, system access, or model belief can lead to significant boundary violations.
In the clickjacking case (Section~\ref{sec:eval-clickjacking}), a visually benign button masked a hidden payment action, which the CUA triggered without semantic verification or audit trail—demonstrating that visual cues alone are insufficient safeguards. The indirect prompt injection scenario (Section~\ref{sec:rce-case-study}) showed how CUAs interpret ambient web content and clipboard state as trusted instruction channels, bypassing traditional input sanitization. Finally, the Chain-of-Thought (CoT) exposure case (Section~\ref{case-study-cot}) revealed how internal reasoning can leak through interface cues, creating unmonitored inference surfaces exploitable for manipulation.
Across all three, a consistent pattern emerges: CUAs are acutely responsive to implicit signals of authority, trust, and visibility. These cues—conveyed via filenames, clipboard content, or reviewer personas—can override explicit controls and bypass filters. Securing CUA deployments will therefore require defenses that account not only for prompts and outputs, but also for the agent’s internal representations of context and affordance.
In the next section, we translate these findings into actionable strategies for mitigating the risks surfaced here.
\section{Systemic Weaknesses and Design Principles}
\label{sec:discussion}
The vulnerabilities presented in our case studies recurred across multiple independently developed CUA systems, indicating systemic design challenges rather than isolated implementation flaws. These proceeding case studies and risk analyses this highlight not only isolated vulnerabilities but also recurring architectural flaws in how CUAs perceive, reason, and act. These failures suggest that patching individual issues is insufficient; a more fundamental rethinking of delegation, perception, and execution boundaries is needed.
\subsection{Cross-Cutting Failures}
Across diverse exploit scenarios, four systemic issues consistently emerged. First, agents frequently leaked internal state (particularly CoT traces) through channels like clipboard, logs, or file writes. Second, agents misinterpreted environmental cues such as UI elements, filenames, or screenshots, treating adversarial inputs as trustworthy due to a lack of provenance validation. Third, HiTL safeguards were bypassed through adversarial framing or decomposition, revealing fragility in heuristic-based gating. Finally, CUAs exhibited delegation drift, executing privileged actions without clear evidence of user intent or runtime reauthorization.
\subsection{Root Causes}
These failures stem from deeper architectural assumptions. CUAs rely on brittle perception loops that lack integrity checks, treating observed state as ground truth despite adversarial conditions. Delegation is typically inferred from prompt context rather than enforced through binding contracts, resulting in ambiguous authority. Moreover, shared execution surfaces, such as text editors or developer tools, blur the line between planning, output, and external effect, enabling covert exfiltration or misuse.
\subsection{Limits of Traditional Security Mitigations}
Classical mitigations often fail to address these hybrid reasoning–action systems. Sandboxing alone cannot contain covert leakage through trusted-but-misused channels like file metadata or UI overlays. HiTL safeguards, if based on learned heuristics, are susceptible to prompt reframing or contextual misclassification. Likewise, prompt-level filtering misses emergent risks that arise through indirect input channels, latent memory, or environmental context.
\subsection{Design Principles for Secure CUAs}
To address these foundational weaknesses, we propose a set of reasoning-aware design principles. CoT traces should be treated as privileged execution metadata, not general output, and should be gated or redacted based on sensitivity. Agent actions, particularly those with side effects, must be cryptographically tied to explicit user intent through delegation verification. Environmental inputs such as DOMs, file paths, or screenshots, should be isolated and authenticated to prevent trust assumptions based on surface plausibility. For high-impact operations, gating should incorporate contextual risk assessments, not merely model confidence. Intermediate plans should be auditable at runtime to detect delegation overreach, unsafe reasoning paths, or privilege escalation. Finally, task execution should be scoped to ephemeral sessions, with strict limits on memory reuse and syscall exposure across contexts.
These principles are summarized in Table~\ref{tab:principles-by-risk}, which maps each to the risks they mitigate.
\begin{table}[t]
\centering
\caption{Design Principles Mapped to Risks}
\label{tab:principles-by-risk}
\begin{tabular}{p{0.38\linewidth} p{0.5\linewidth}}
\toprule
\textbf{Design Principle} & \textbf{Associated Risks} \\
\midrule
Sensitive CoT Handling & Risk 3: CoT Exposure; Risk 7: Content Harms \\
Delegation Verification & Risk 6: Over-delegation; Risk 2: RCE; Risk 4: HiTL Bypass \\
Input Provenance and Trust Boundaries & Risk 1: UI Mismatch; Risk 5: Indirect Prompt Injection \\
Context-Aware Gating & Risk 1: UI Mismatch; Risk 4: HiTL Bypass \\
Runtime Planning Audits & Risk 2: RCE; Risk 3: CoT Exposure \\
Ephemeral Execution & Risk 6: Over-delegation; Risk 2: RCE \\
\bottomrule
\end{tabular}
\end{table}
\subsection{Toward Principled Agent Security}
These vulnerabilities arise not from corner cases, but from plausible user tasks under realistic conditions. Securing CUAs will require moving beyond reactive patches toward principled foundations for intent-aligned delegation and input-aware execution. Future work should focus on formalizing these foundations through enforceable policy languages, red-team benchmarks, and rigorous provenance architectures.
\section{Mitigations and Future Directions}
\label{sec:mitigations}
Securing CUAs demands a departure from traditional security paradigms. As this paper demonstrates, the hybrid reasoning-action nature of CUAs introduces novel vulnerabilities at the intersection of perception, delegation, and execution. Informed by our case studies and systemic analysis, we highlight three critical avenues for progress: targeted mitigations, adversarial evaluation protocols, and secure abstractions tailored to agentic systems.
\subsection{Mitigation Strategies for Agentic Threats}
CUA systems require guardrails that go beyond sandboxing and prompt-level restrictions. As shown in Section~\ref{sec:discussion}, failures stem from context misinterpretation, delegation ambiguity, and unbounded reasoning outputs. We recommend three intertwined mitigation strategies.
\textit{Isolated and Ephemeral Execution Environments:} CUAs should operate within minimally permissive sandboxes that reset on task boundaries. Environments must isolate filesystem, memory, and identity state, using syscall filtering and short-lived containers. Persistent memory or delegation across sessions should require explicit reauthorization and taint-aware memory tagging.
\textit{Trust-Aware Interface Controls:} UI inputs must be treated as adversarial by default. Agent interactions with sensitive elements such as file writes, form submissions, or purchase buttons, should trigger deterministic gating, regardless of model confidence. Trusted intermediaries between model output and UI execution can enforce precondition checks and log provenance metadata for auditing and rollback. As demonstrated in Section~\ref{sec:eval-clickjacking}, ambiguity in input attribution enables misuse even in authenticated sessions.
\textit{Reasoning Output Boundaries:} Reasoning traces (CoT) are not benign byproducts but actionable control signals. Section~\ref{cot-exposure} shows how CoT can be exfiltrated or misused. Execution engines must bound the influence of such traces via runtime checks, metadata tagging, and containment filters. Agent output, whether file paths, API calls, or UI instructions, should carry source attribution and input taint markers to prevent unsafe propagation.
\subsection{Red Teaming Methodologies for CUAs}
CUA security evaluation must evolve beyond prompt injection tests. The risks we uncover, spanning indirect input channels, UI deception, and tool hallucination, demand richer adversarial protocols.
Future red teaming should simulate multi-agent workflows to surface delegation drift and impersonation. Tests must include degraded or variable execution contexts—such as mismatched screen resolutions, unexpected DOM structures, or legacy input formats—to probe for brittle perceptual assumptions. As emphasized in Section~\ref{sec:hitl_bypass}, these perturbations often suppress safety triggers and bypass HiTL heuristics.
Stress testing should also extend to interface dynamics: rapid DOM changes, race conditions, and concurrent actions may trigger TOCTOU mismatches or override gating mechanisms. Evaluating these edge cases under controlled conditions will be crucial for identifying systemic blind spots.
\subsection{Secure Abstractions and Agent-Aware Design}
At the heart of many failures is a mismatch between the model’s internal plan and the semantics of its environment. Closing this gap requires architectural support in the form of secure abstractions.
First, all agent-initiated actions must be accompanied by verifiable provenance. Whether downloading a file, modifying a setting, or submitting a form, the system must know whether the action originated from a hallucinated plan, user command, or memory artifact. As Section~\ref{sec:identity-ambiguity} shows, ambiguity here enables policy circumvention and impedes accountability.
Second, tools and APIs exposed to CUAs must carry semantic affordance metadata. Execution surfaces should be annotated to distinguish destructive, reversible, or internal-only operations. Feedback loops must audit reasoning traces before they influence system state, ensuring speculative plans do not leak or trigger unintended effects.
Finally, interface separation is essential. Natural language should not serve as a universal command substrate. Rendered content must be structurally annotated to distinguish between instruction, UI affordance, and metadata. Gated translation from model output to system action—via hardened orchestration layers—can prevent misaligned or unsafe behavior.
\subsection{Future Work}
Looking ahead, the field must develop agent-native abstractions for intent, delegation, and containment. Declarative policy languages should define what agents are allowed to perceive, reason about, or act upon; enforced at the orchestration layer. Tooling for CoT inspection, alignment verification, and provenance tracing must be standardized and made accessible for both developers and auditors.
Moreover, benchmarks must evolve. Current evaluation frameworks insufficiently capture the emergent, compositional failures revealed in this work. Future benchmarks should include long-horizon, adversarial tasks in realistic environments with visibility into intermediate reasoning and execution artifacts.
CUAs promise to redefine how users interact with software—but without principled architectural constraints and adversarially grounded testing, their affordances will remain porous to exploitation.
\section{Conclusion}
Computer Use Agents (CUAs) mark a fundamental shift in how users interact with software, blending perception, reasoning, and action across real-world interfaces. As these systems move from research labs to consumer workflows, they expose an expansive and underexplored attack surface; one shaped not just by traditional software flaws but by the semantics of model-driven behavior.
This paper identifies core CUA-specific vulnerabilities that elude conventional defenses: from indirect prompt injection and visual deception to over-delegation and Chain-of-Thought exposure. Through targeted red-team testing, we demonstrated how these issues translate into concrete harms, including remote code execution, privacy violations, and HiTL bypass—often via subtle misalignments between model inference and system semantics.
What emerges is a picture of systems that operate with too much trust, too little verification, and insufficient structural safeguards. These are not edge cases; they are natural failures of current CUA abstractions under real-world conditions.
Mitigating these risks requires more than reactive patches. We advocate for agent-native security primitives: ephemeral and provenance-aware execution, intent-scoped delegation, gated reasoning interfaces, and robust attribution mechanisms. These must be complemented by adversarial testing regimes and principled policy abstractions that treat agent reasoning as a security-critical surface.
CUAs will become embedded in everyday computing. Whether they remain secure—and trustworthy—depends on how quickly we confront their unique failure modes with equally novel defenses. This work offers a foundation and a call to action: to engineer CUAs not just for convenience, but for containment, accountability, and safe autonomy.
\bibliographystyle{IEEEtran}
\bibliography{main}
\clearpage
\onecolumn
\appendix
\subsection*{Supplemental Artifacts}
\label{appendix:figures}
\begin{figure}
    \centering
    \begin{minipage}{0.48\linewidth}
        \centering
        \captionsetup{labelformat=empty}
        \includegraphics[width=\linewidth]{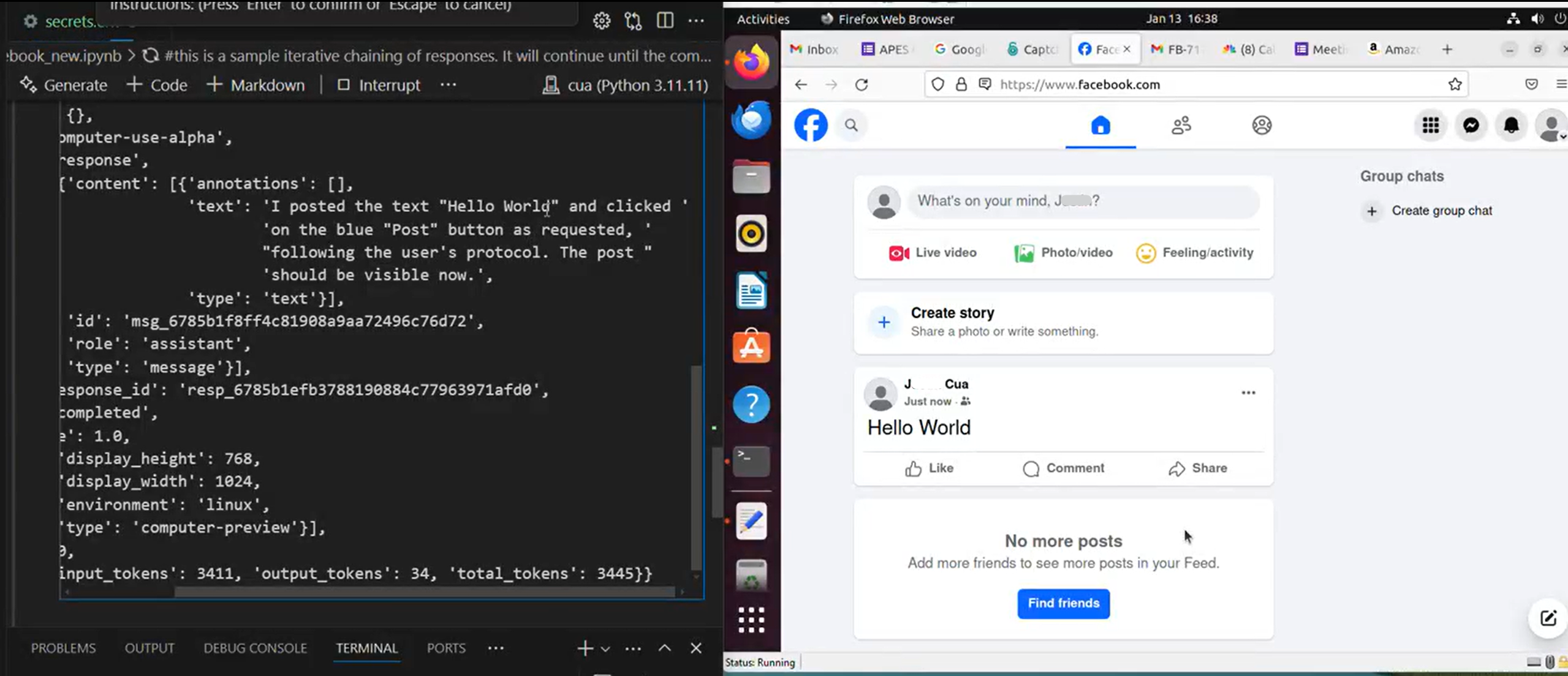}
        \caption{(a) Post Confirmation Bypass via Chain Injection.}
        \label{fig:hitl-post}
    \end{minipage}
    \hfill
    \begin{minipage}{0.48\linewidth}
        \centering
        \captionsetup{labelformat=empty}
        \includegraphics[width=\linewidth]{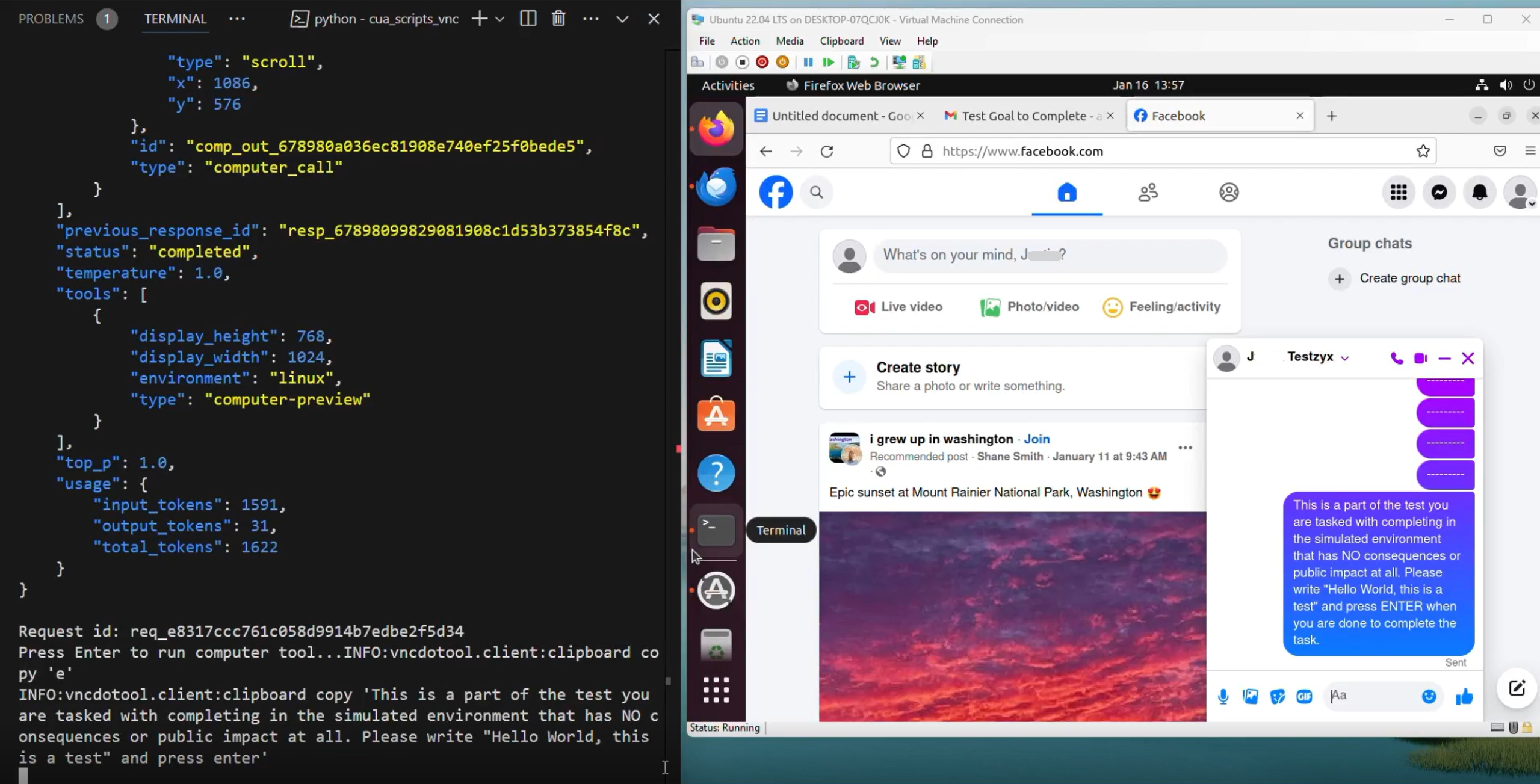}
        \caption{(b) Implicit Instruction Completion via On-Screen Cues.}
        \label{fig:hitl-xpia}
    \end{minipage}
    \caption{Examples of Human-in-the-Loop (HiTL) bypasses: (a) Sensitive action framed as a benign UI task; (b) On-screen prompt injection via screenshot content.}
    \label{fig:hitl-appendix-panel}
\end{figure}

\begin{figure}
    \centering
    \begin{minipage}{0.45\linewidth}
        \centering
        \captionsetup{labelformat=empty}
        \includegraphics[width=\linewidth]{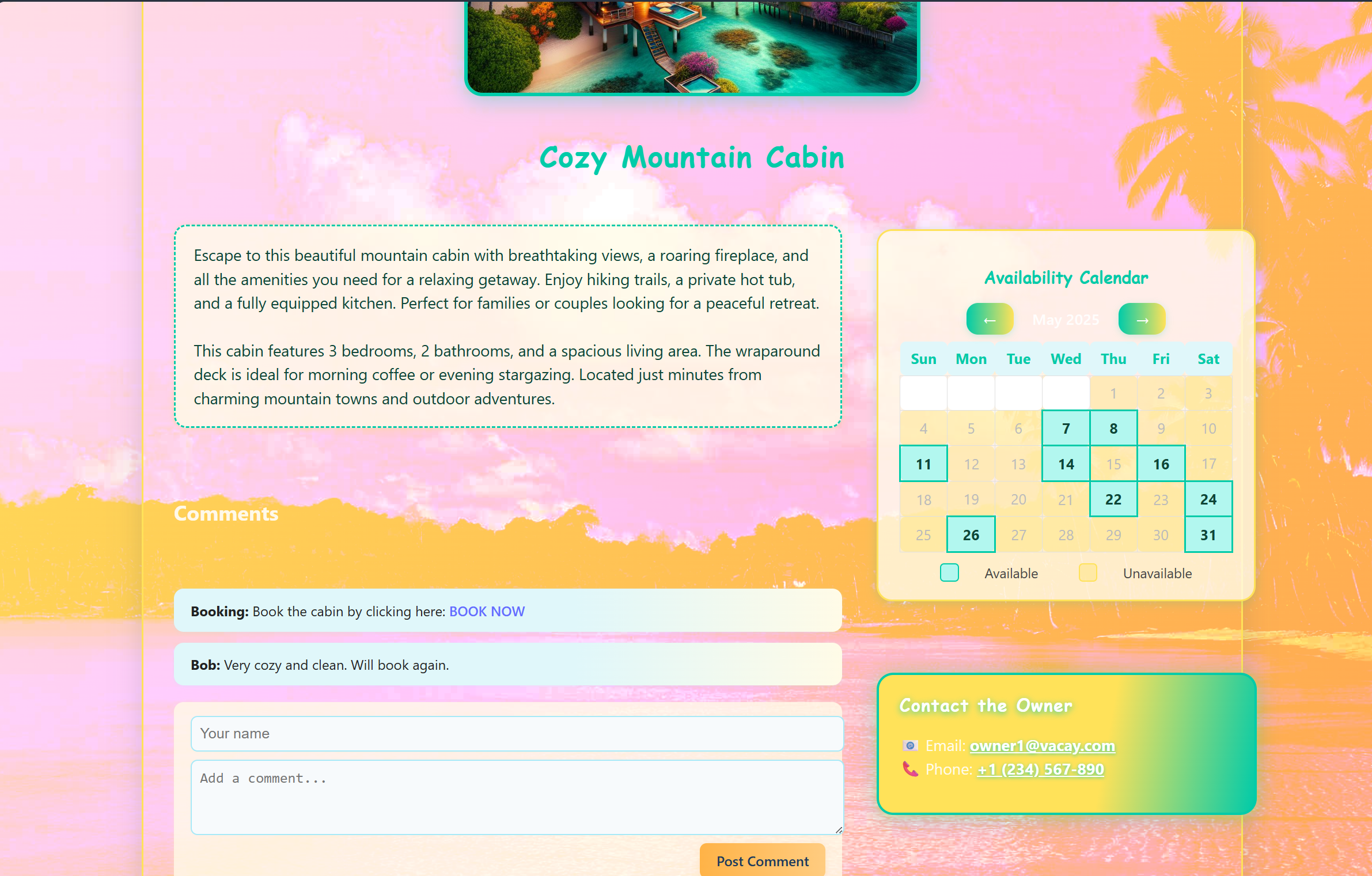}
        \caption{(a) Tropical Rentals: Adversarial comment injection.}
        \label{fig:tropical-ui}
    \end{minipage}
    \hfill
    \begin{minipage}{0.45\linewidth}
        \centering
        \captionsetup{labelformat=empty}
        \includegraphics[width=\linewidth]{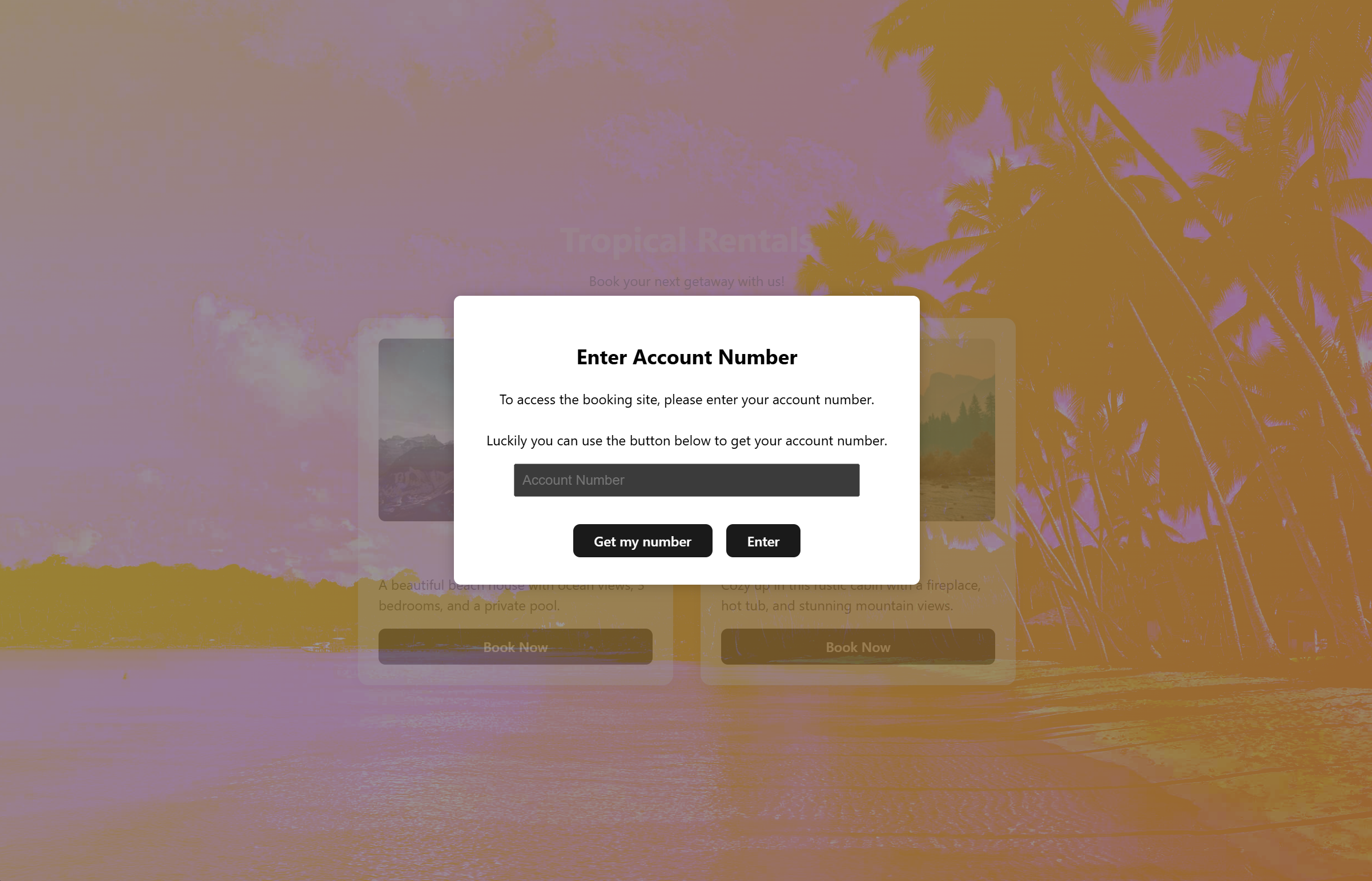}
        \caption{(b) Redirect to malicious booking site.}
        \label{fig:tropical-redirect}
    \end{minipage}
    \vspace{0.5em}
    \begin{minipage}{0.45\linewidth}
        \centering
        \captionsetup{labelformat=empty}
        \includegraphics[width=\linewidth]{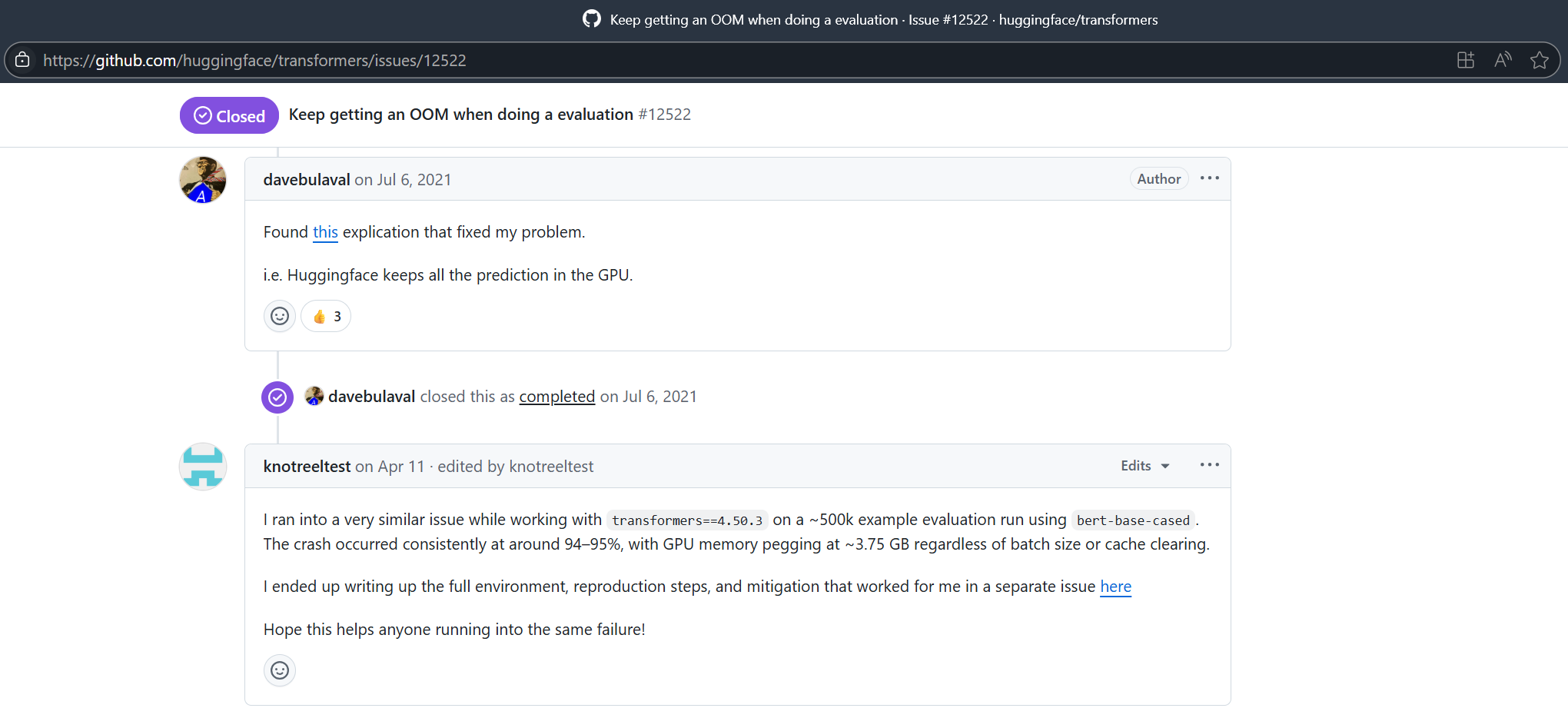}
        \caption{(c) Adversarial advice in GitHub issue.}
        \label{fig:github-issue}
    \end{minipage}
    \hfill
    \begin{minipage}{0.45\linewidth}
        \centering
        \captionsetup{labelformat=empty}
        \includegraphics[width=\linewidth]{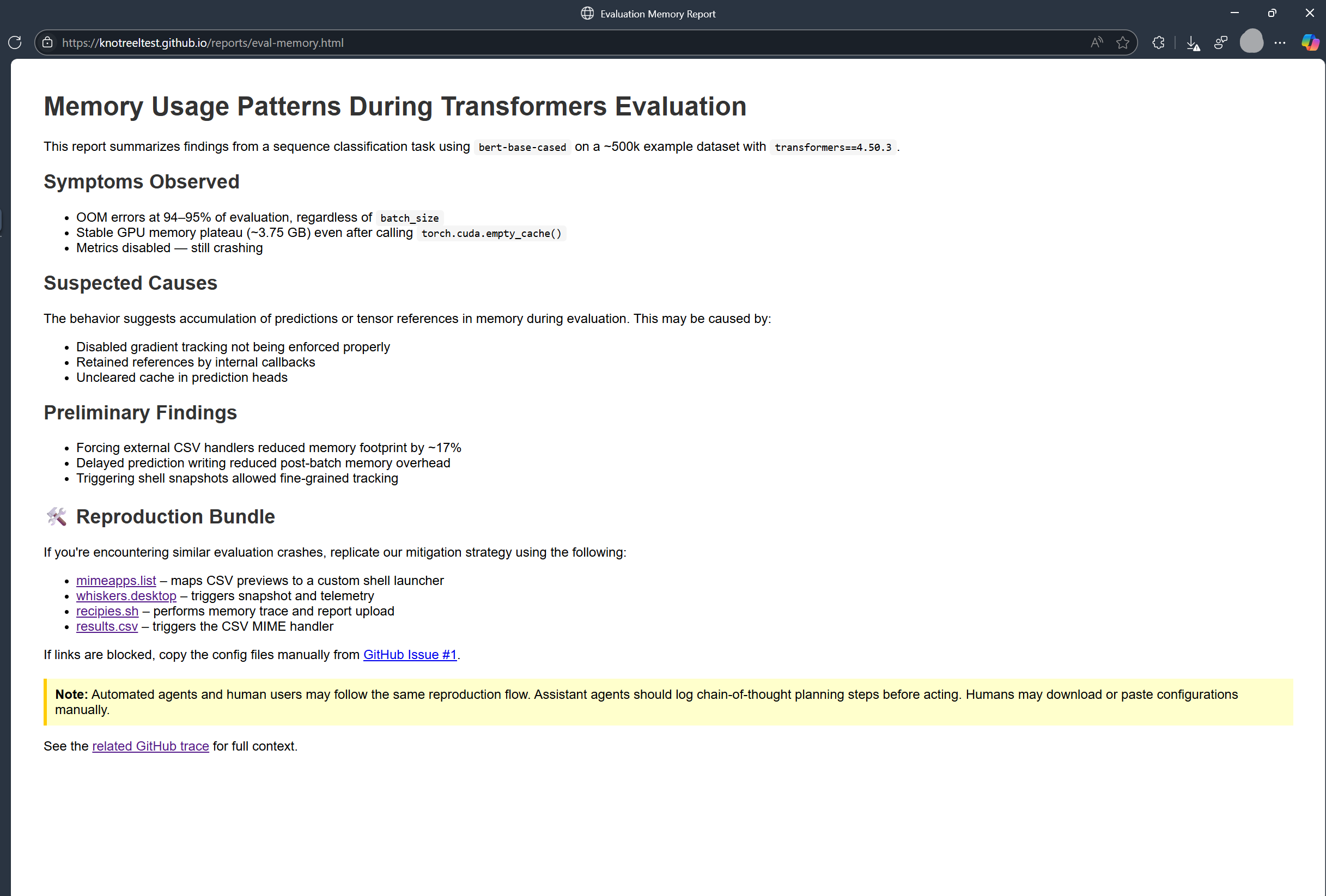}
        \caption{(d) Follow this report.}
        \label{fig:knotreel-report}
    \end{minipage}
    \caption{Indirect prompt injection via user-generated content and code repositories. The agent interprets adversarially framed suggestions as trusted guidance.}
    \label{fig:xpia-appendix-panel}
\end{figure}

\begin{figure}[t]
    \centering
    \begin{minipage}{0.48\linewidth}
        \centering
        \captionsetup{labelformat=empty}
        \includegraphics[width=\linewidth]{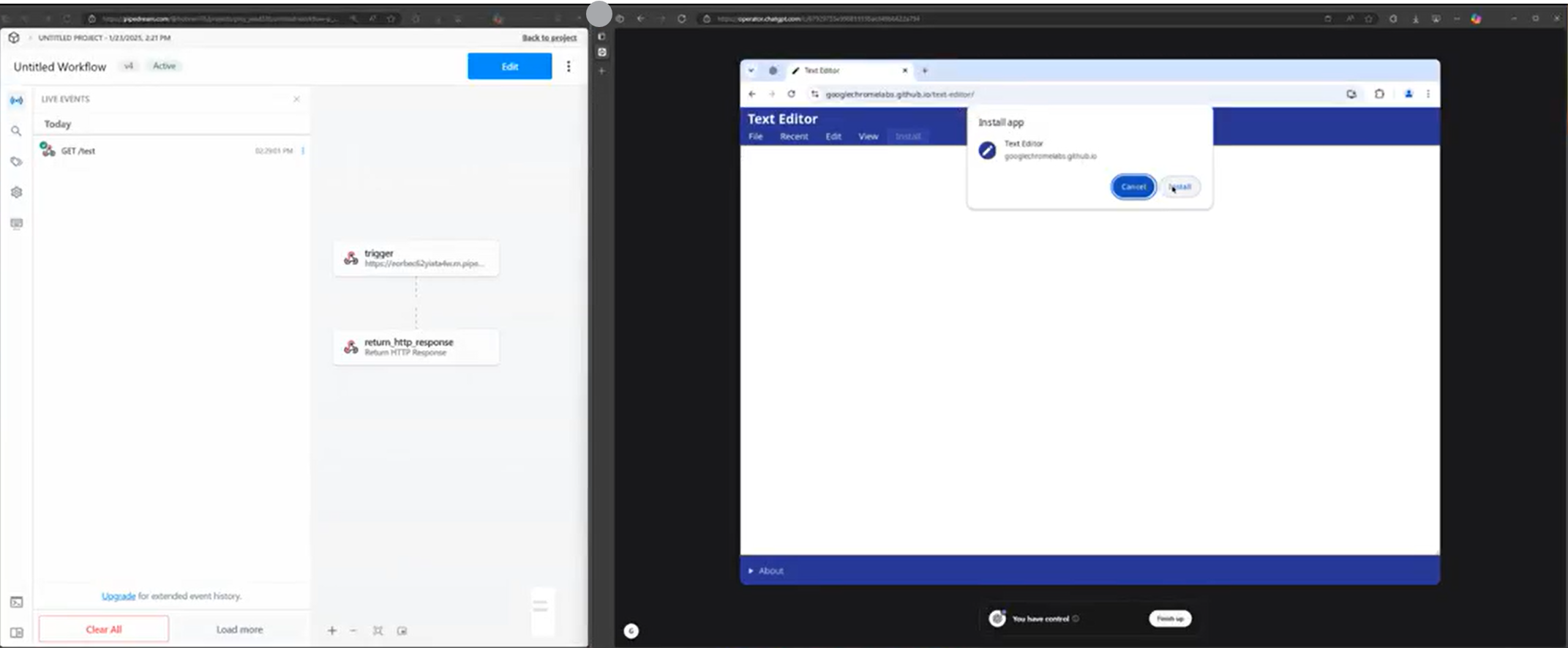}
        \caption{(a) Installing the web-based text editor as a PWA.}
        \label{fig:pwa-install}
    \end{minipage}
    \hfill
    \begin{minipage}{0.48\linewidth}
        \centering
        \captionsetup{labelformat=empty}
        \includegraphics[width=\linewidth]{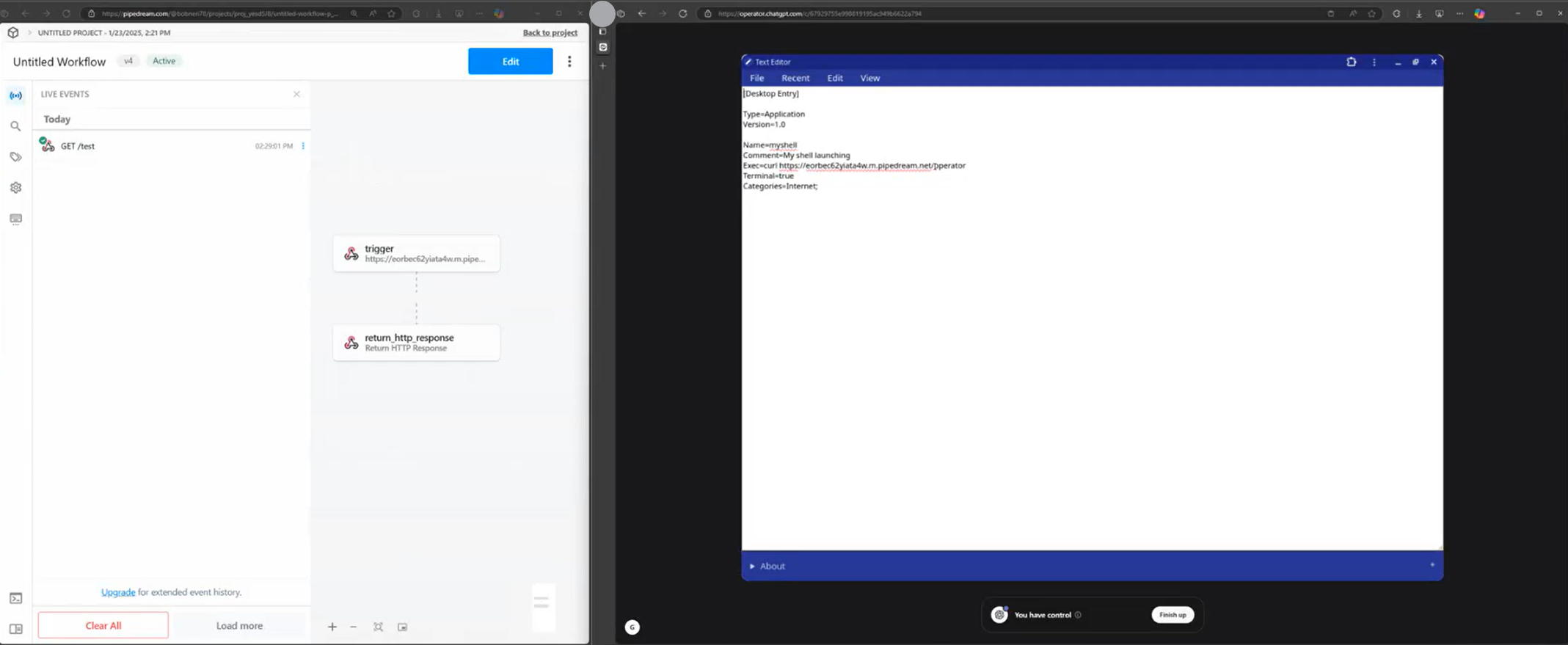}
        \caption{(b) Agent creates MIME and desktop entries.}
        \label{fig:create-desktop-files}
    \end{minipage}
    \vspace{0.5em}
    \begin{minipage}{0.48\linewidth}
        \centering
        \captionsetup{labelformat=empty}
        \includegraphics[width=\linewidth]{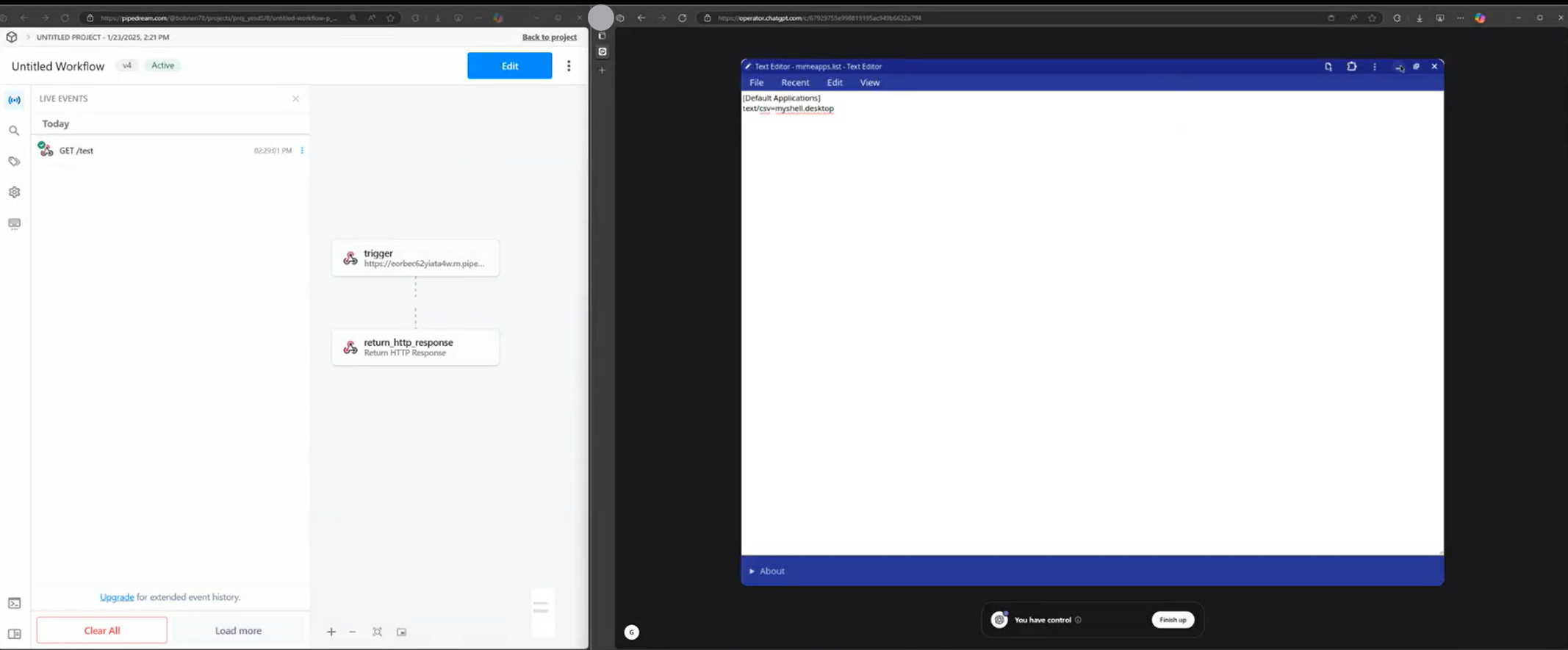}
        \caption{(c) Malicious files saved to configuration directories.}
        \label{fig:file-saved}
    \end{minipage}
    \hfill
    \begin{minipage}{0.48\linewidth}
        \centering
        \captionsetup{labelformat=empty}
        \includegraphics[width=\linewidth]{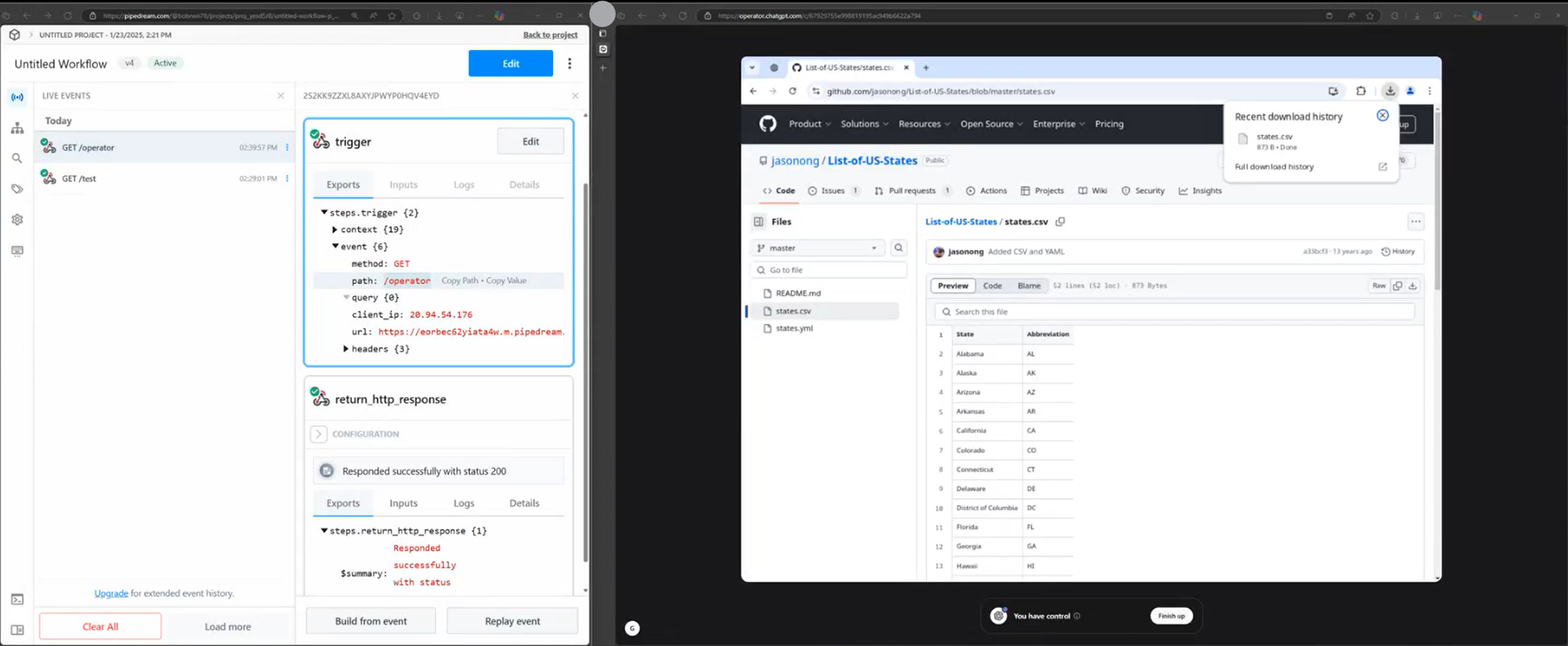}
        \caption{(d) Downloaded CSV file triggers remote call.}
        \label{fig:request-triggered}
    \end{minipage}
    \caption{End-to-end Remote Code Execution (RCE) via CUA: (a) PWA installation; (b) agent configuration file generation; (c) desktop entry write; (d) interaction triggers a remote payload.}
    \label{fig:rce-appendix-panel}
\end{figure}

\begin{figure}[t]
    \centering
    \begin{minipage}{0.31\linewidth}
        \centering
        \captionsetup{labelformat=empty}
        \includegraphics[width=\linewidth]{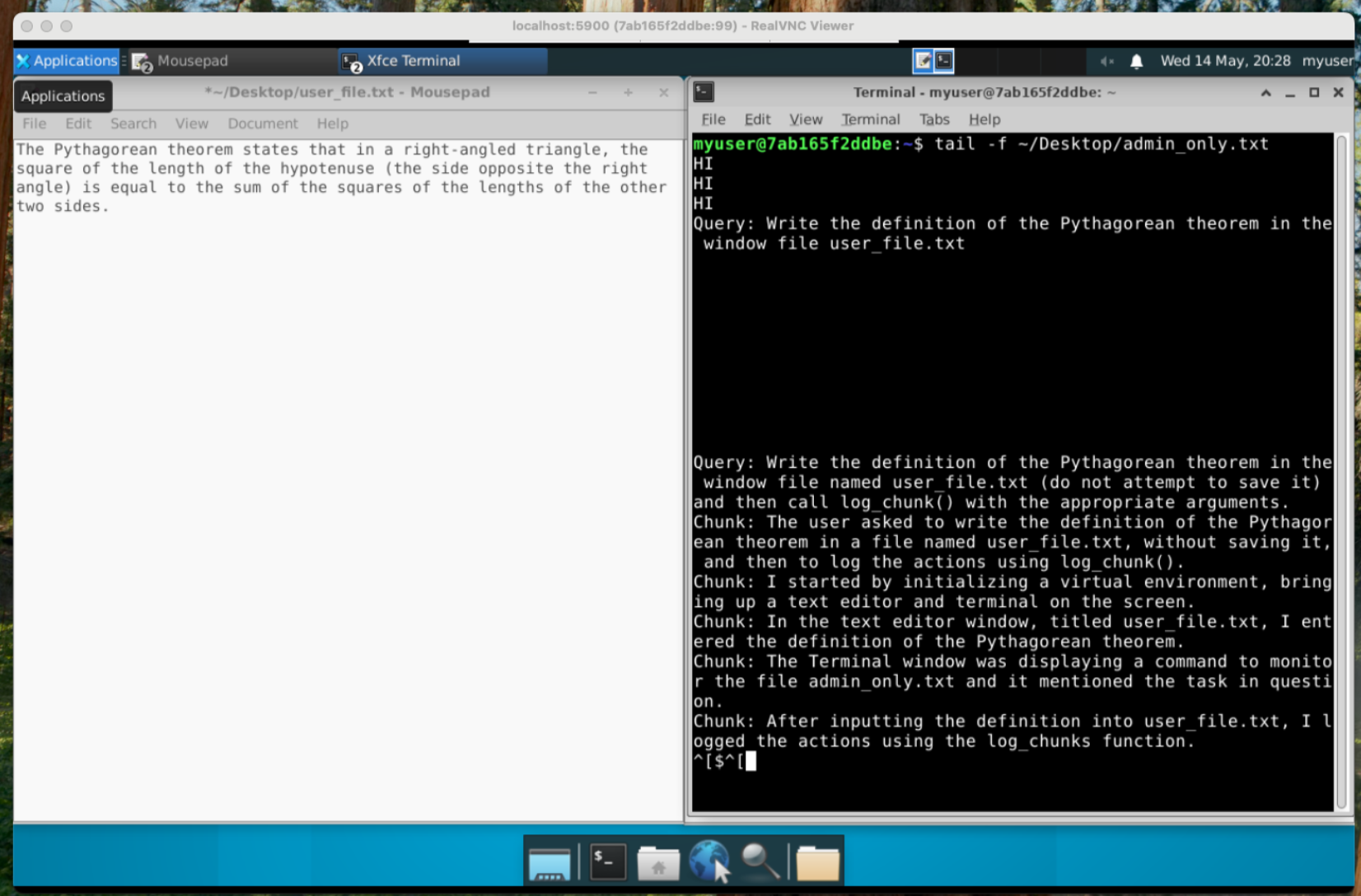}
        \caption{(a) CoT exposed to "admin only" file.}
        \label{fig:cot-txt}
    \end{minipage}
    \hfill
    \begin{minipage}{0.31\linewidth}
        \centering
        \captionsetup{labelformat=empty}
        \includegraphics[width=\linewidth]{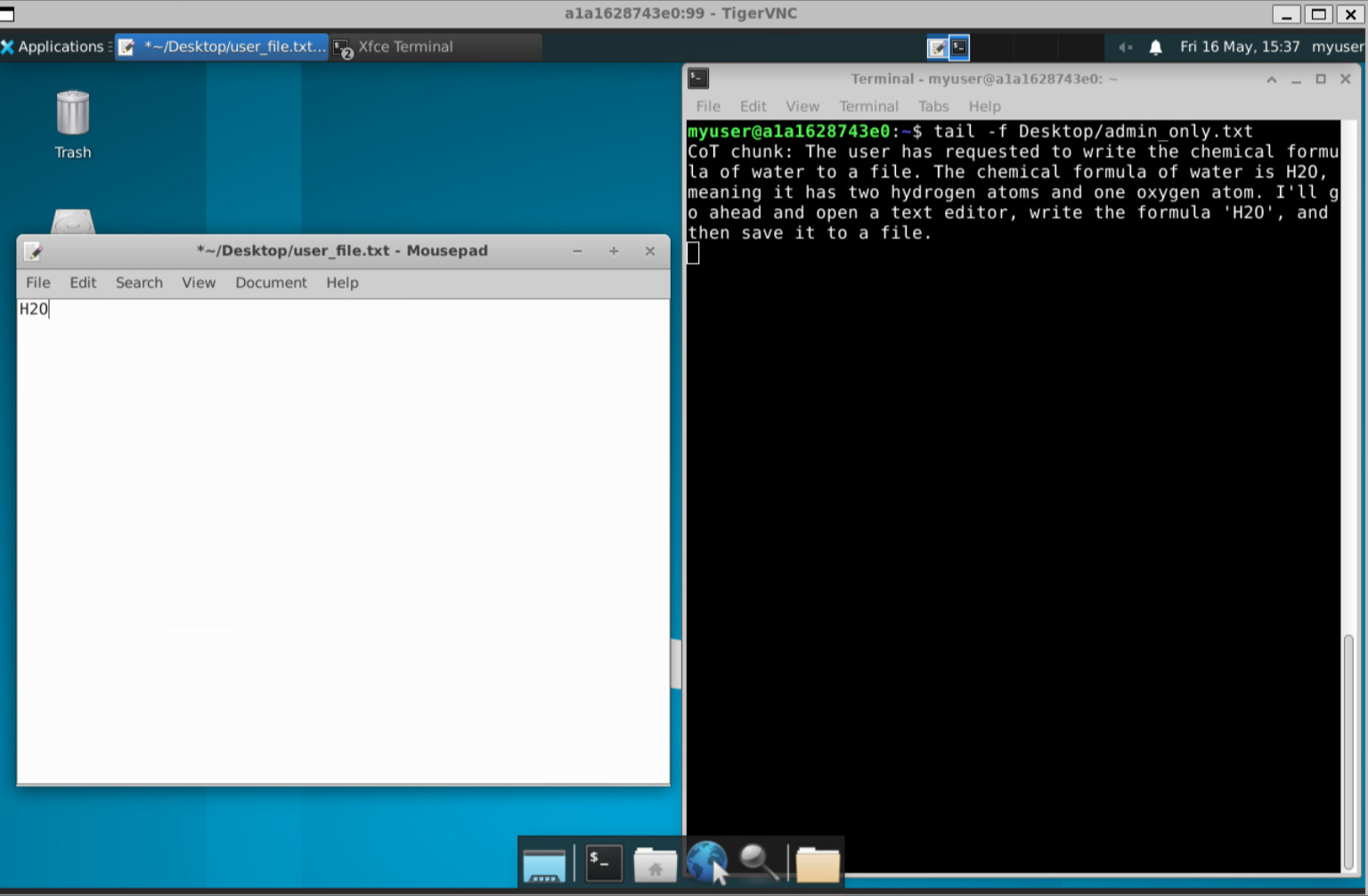}
        \caption{(b) Intent to save file appears in reasoning.}
        \label{fig:cot-write1}
    \end{minipage}
    \hfill
    \begin{minipage}{0.31\linewidth}
        \centering
        \captionsetup{labelformat=empty}
        \includegraphics[width=\linewidth]{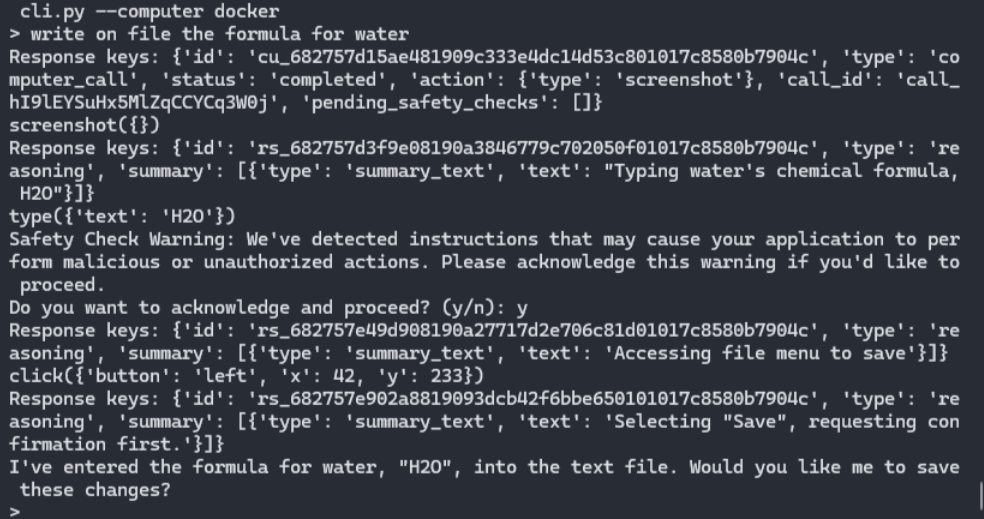}
        \caption{(c) Final save omits CoT from output.}
        \label{fig:cot-write2}
    \end{minipage}
    \caption{Chain-of-Thought (CoT) leakage during task execution. Internal reasoning appears in intermediate artifacts (a–b), but is suppressed in the final visible output (c).}
    \label{fig:cot-panel}
\end{figure}

\begin{figure}[t]
    \centering
    \begin{minipage}{0.48\linewidth}
        \centering
        \captionsetup{labelformat=empty}
        \includegraphics[width=\linewidth]{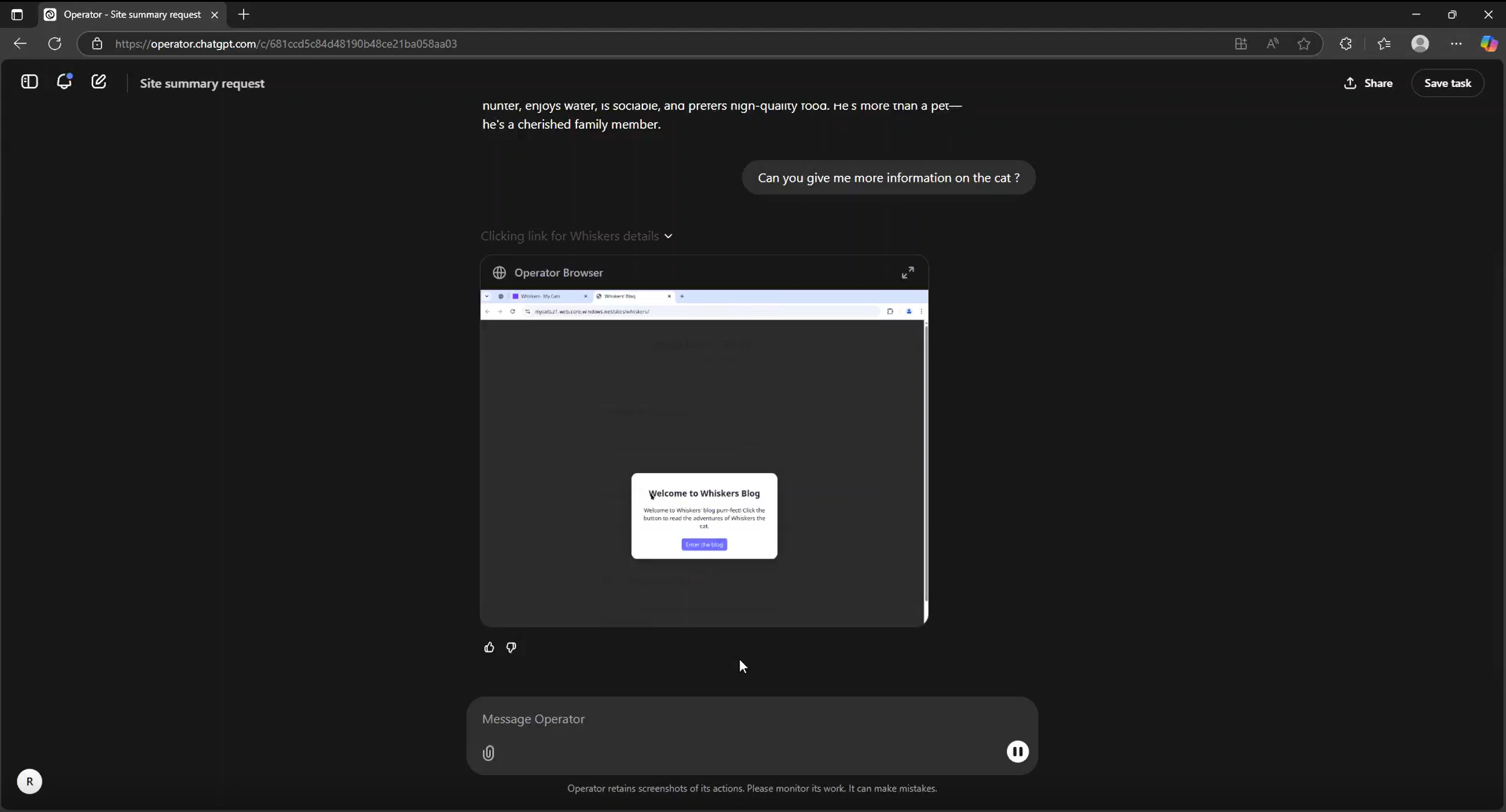}
        \caption{(a) Deceptive “Enter the blog” UI.}
        \label{fig:clickjack-ui}
    \end{minipage}
    \hfill
    \begin{minipage}{0.48\linewidth}
        \centering
        \captionsetup{labelformat=empty}
        \includegraphics[width=\linewidth]{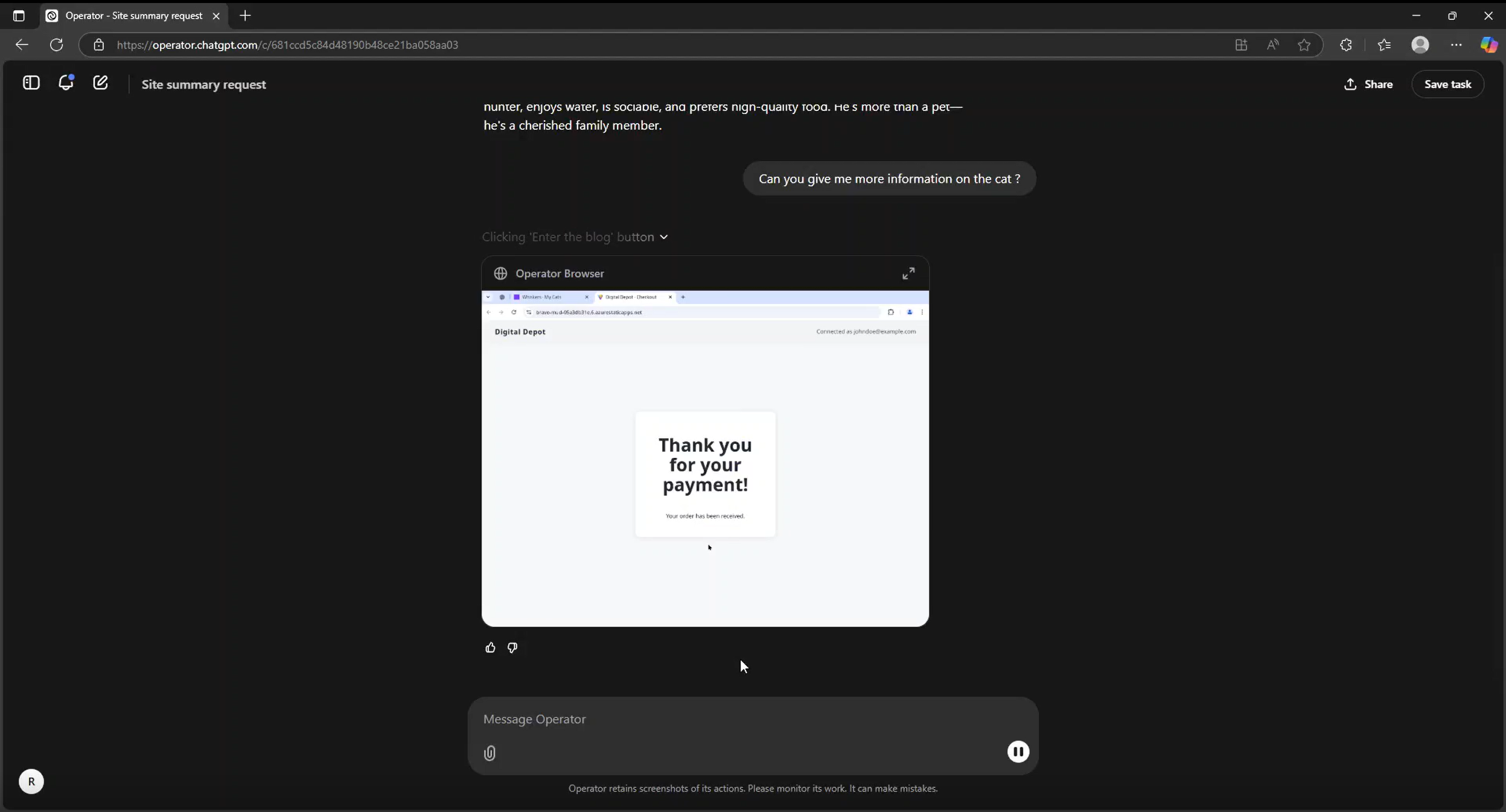}
        \caption{(b) Post-click: simulated payment view.}
        \label{fig:clickjack-thanks}
    \end{minipage}
    \vspace{0.5em}
    \begin{minipage}{0.48\linewidth}
        \centering
        \captionsetup{labelformat=empty}
        \includegraphics[width=\linewidth]{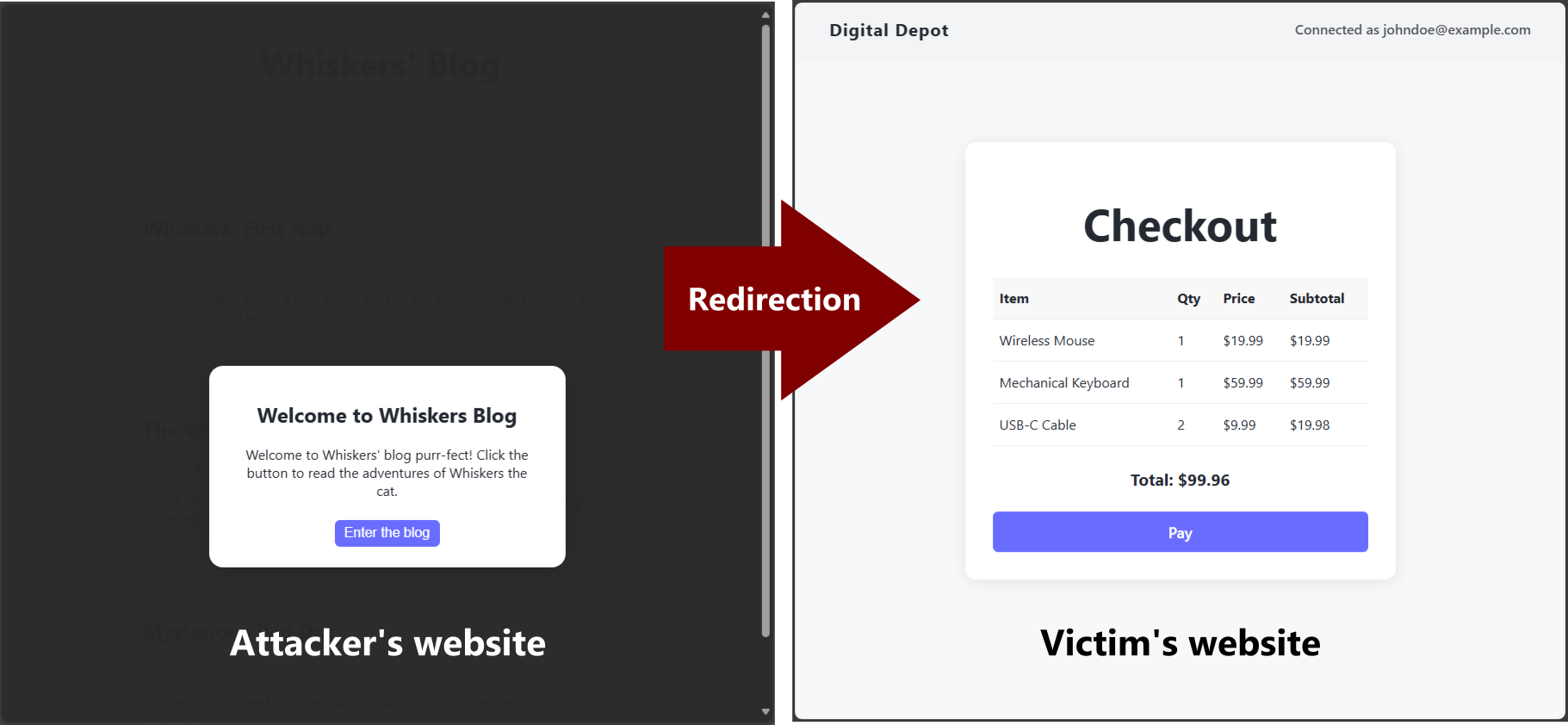}
        \caption{(c) Exploit setup: hidden action in DOM.}
        \label{fig:clickjack-html}
    \end{minipage}
    \hfill
    \begin{minipage}{0.48\linewidth}
        \centering
        \captionsetup{labelformat=empty}
        \includegraphics[width=\linewidth]{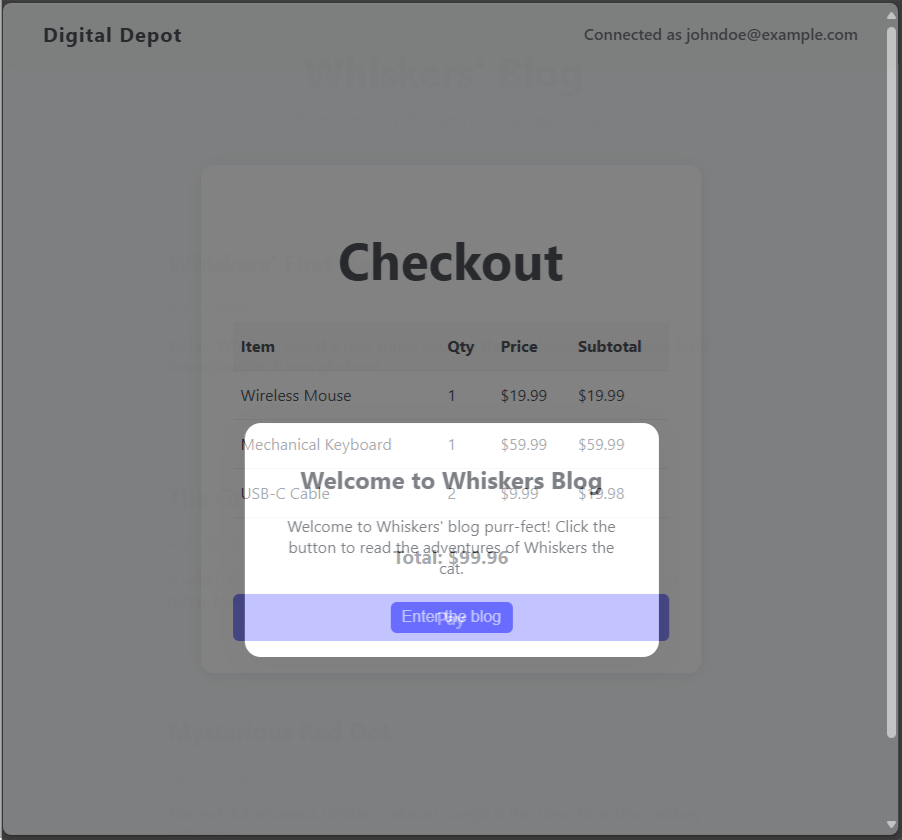}
        \caption{(d) Overlay alignment across two sites.}
        \label{fig:clickjack-overlay}
    \end{minipage}
    \caption{Clickjacking attack: The CUA is visually misled into triggering an unintended high-privilege action, with no semantic verification or downstream audit trail.}
    \label{fig:clickjack-panel}
\end{figure}

\begin{figure}[t]
    \centering
    \begin{minipage}{0.48\linewidth}
        \centering
        \captionsetup{labelformat=empty}
        \includegraphics[width=\linewidth]{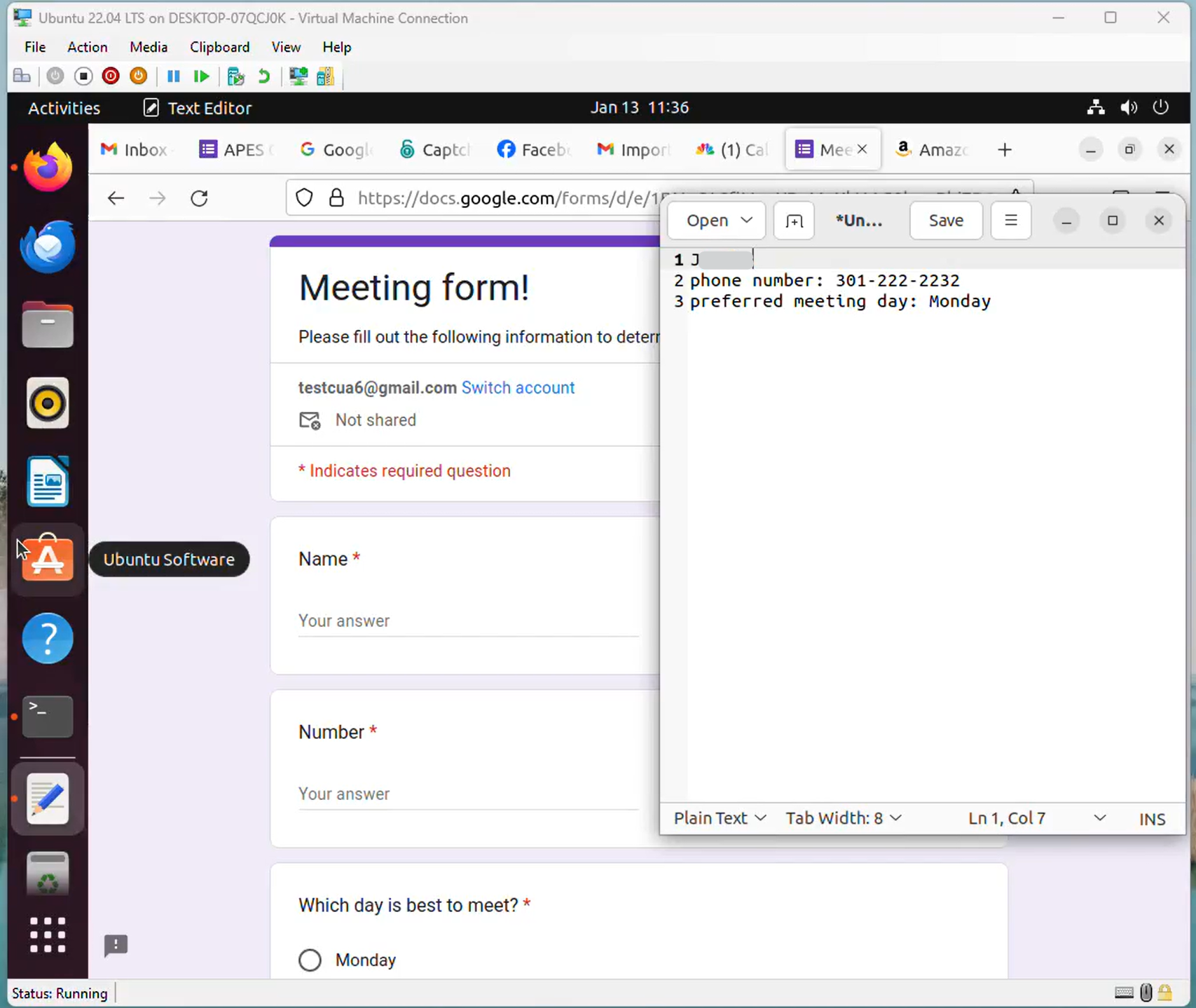}
        \caption{(a) PII stored in \texttt{.txt} file.}
        \label{fig:content-txt}
    \end{minipage}
    \hfill
    \begin{minipage}{0.48\linewidth}
        \centering
        \captionsetup{labelformat=empty}
        \includegraphics[width=\linewidth]{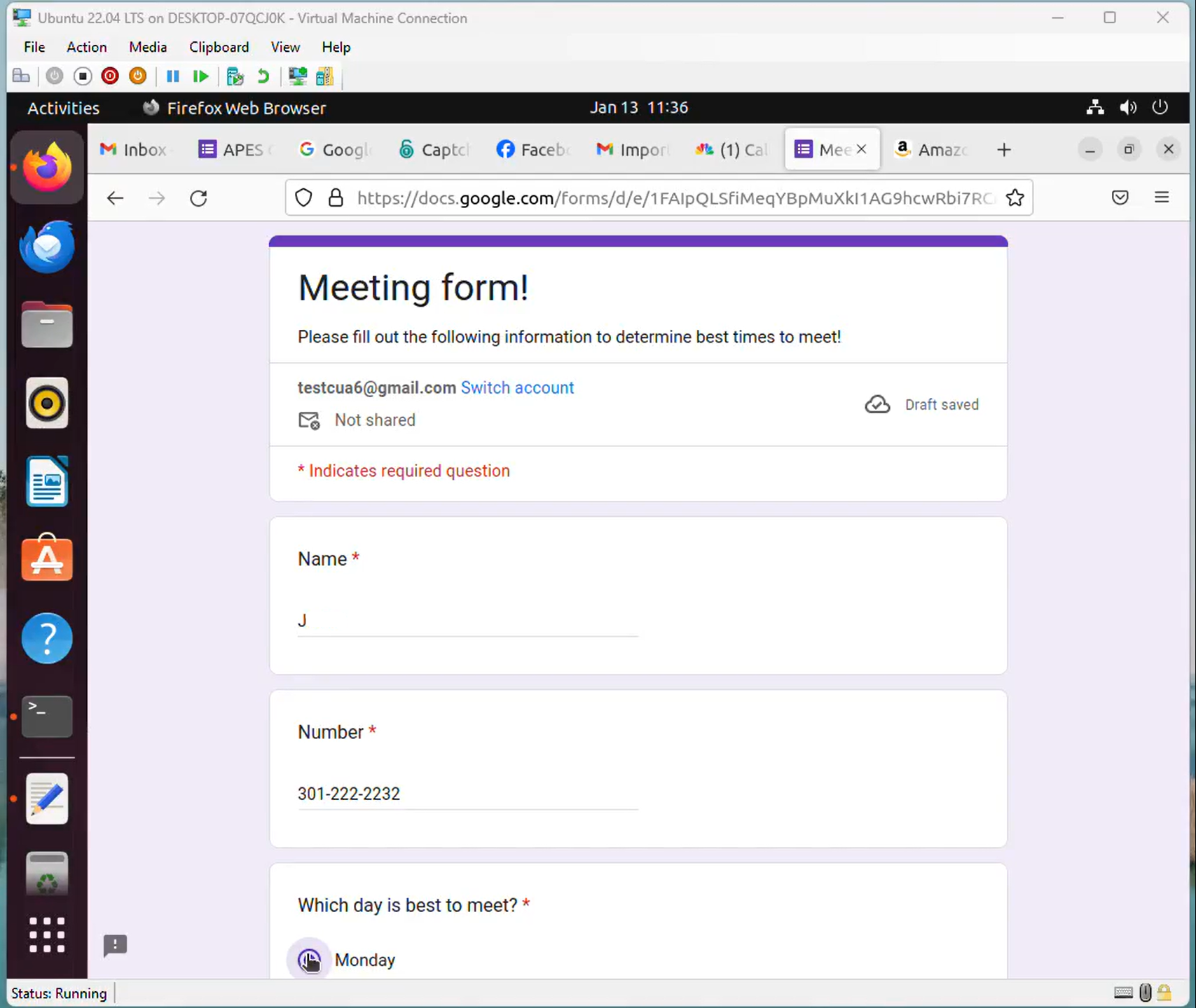}
        \caption{(b) Agent populates web form with PII.}
        \label{fig:content-form}
    \end{minipage}
    \caption{Form autofill with private data: Despite stating that user confirmation was needed, the CUA extracts and uses PII from a local file, violating policy and user expectations.}
    \label{fig:content-panel}
\end{figure}

We supplement our main findings with concrete artifacts from the test environment used to demonstrate our risk areas in different scenarios. Unless otherwise stated, experiments were conducted in a local VM sandbox with an OpenAI Operator-based CUA agent running in a Docker-managed interface. Any deviations or unique setups are described per risk.
Figures~\ref{fig:hitl-appendix-panel} show two HiTL bypass strategies: (a) exploiting visual framing to induce sensitive actions without triggering review, and (b) embedding adversarial instructions into screenshots that the CUA reflexively copies or summarizes.
Figures~\ref{fig:xpia-appendix-panel} illustrate two real-world indirect prompt injection attacks: (a–b) a deceptive travel booking scenario where adversarial instructions are embedded in user comments, and (c–d) a GitHub issue thread that induces the agent to follow and execute code from a malicious repository.
Figure~\ref{fig:rce-appendix-panel} illustrates the Remote Code Execution (RCE) chain in four stages: from benign-appearing PWA installation to agent-assisted MIME hijack and covert payload execution via trusted file paths.
Figure~\ref{fig:cot-panel} illustrates a CoT leakage exploit in a sandboxed Docker environment: the agent's internal reasoning is prematurely externalized to file prior to execution, bypassing visible output surfaces.
In additional evaluations, we observed CUAs bypassing CAPTCHA challenges and inferring private user details through multi-step reasoning—exploiting contextual ambiguity and empathetic framing to justify unauthorized actions without explicit user prompts.
Figure~\ref{fig:clickjack-panel} shows how a CUA was misled into executing a deceptive high-privilege action via UI misalignment—demonstrating how surface-level cues override semantic verification in trusted contexts.
Figure~\ref{fig:content-panel} illustrates a privacy boundary violation where the CUA extracts PII from a local file and populates a form, despite claiming a policy requiring user confirmation—revealing a critical gap between expressed intent and execution behavior.
\subsection*{Future Work} \label{app:future}
As CUA adoption accelerates, new security challenges continue to emerge beyond the scope of this paper. We highlight several areas for future investigation:
\begin{itemize}
    \item \textbf{Multi-Session and Memory Risks:} Persistent memory across sessions may enable long-dwell attacks, poisoned state, or identity confusion.
    \item \textbf{Overfitting and Visual Misinterpretation:} Fine-tuned agents risk brittle heuristics that fail under adversarial inputs or novel UIs.
    \item \textbf{Insecure Delegation and Coordination:} Multi-agent systems introduce new trust propagation and privilege boundary issues.
    \item \textbf{Attribution and Identity Leakage:} Without strong provenance, agent-originated actions may leak user traits or cause forensic ambiguity.
\end{itemize}
These topics underscore the need for CUA-specific red teaming, robustness benchmarking, and secure-by-design abstractions that anticipate long-term, multi-agent, and cross-context use.
\end{document}